\documentclass[3p]
{sarticle}
\newcommand{\be}{\begin{equation}}
\newcommand{\ee}{\end{equation}}
\newcommand{\beq}{\begin{eqnarray}}
\newcommand{\eeq}{\end{eqnarray}}
\newcommand{\lb}[1]{\label{#1}}

\newcommand{\dd}{{\mathrm{d}}}
\newcommand{\up}[1]{\raisebox{0.7mm}{$\scriptstyle \, #1 $}}
\newcommand{\dn}[1]{\raisebox{-0.7mm}{$\scriptstyle #1 $}}
\usepackage{amsmath}
\usepackage{txfonts}
\usepackage{epsfig}
\usepackage[obeyspaces,hyphens]{url}
\usepackage{graphicx}
\begin{document}
\title{Tsallis statistics in the income distribution of Brazil}

\author[cnen]{Abner D.\ Soares}

\author[ibge]{Newton J.\ Moura Jr.}

\author[if,ov]{Marcelo B.\ Ribeiro\corref{cor1}}
\ead{mbr@if.ufrj.br}

\cortext[cor1]{Corresponding author}
\address[cnen]{Comiss\~ao Nacional de Energia Nuclear -- CNEN,
               Rio de Janeiro, Brazil}
\address[ibge]{Instituto Brasileiro de Geografia e
               Estat\'{\i}stica -- IBGE, Rio de Janeiro, Brazil}
\address[if]{Instituto de F\'{\i}sica, Universidade Federal do
             Rio de Janeiro -- UFRJ, Rio de Janeiro, Brazil}
\address[ov]{Observat\'{o}rio do Valongo, Universidade Federal do
             Rio de Janeiro -- UFRJ, Rio de Janeiro, Brazil}

\begin{abstract}
This paper discusses the empirical evidence of Tsallis
statistical functions in the personal income distribution
of Brazil. Yearly samples from 1978 to 2014 were linearized
by the $q$-logarithm and straight lines were fitted to the
entire range of the income data in all samples, producing a
two-parameters-only single function representation of
the whole distribution in every year. The results showed that
the time evolution of the parameters is periodic and plotting
one in terms of the other reveals a cycle mostly clockwise. It
was also found that the empirical data oscillate periodically
around the fitted straight lines with the amplitude growing as
the income values increase. Since the entire income data range
can be fitted by a single function, this raises questions on
previous results claiming that the income distribution is
constituted by a well defined two-classes-base income structure,
since such a division in two very distinct income classes might
not be an intrinsic property of societies, but a consequence of
an \textit{a priori} fitting-choice procedure that may leave
aside possibly important income dynamics at the intermediate
levels.
\end{abstract}

\begin{keyword}
Income distribution; Brazil's income data; Tsallis statistics;
$q$-logarithm; $q$-exponential
\end{keyword}
\maketitle
\biboptions{sort&compress}
\section{Introduction}\lb{intro}

The functional characterization of the income distribution in a
population is an old problem in economics. Vilfredo Pareto (1848-1923),
widely acknowledged as the first to have studied this problem
systematically, concluded that the richest individuals in a society
have their complementary cumulative income distribution function (see
definition below) behaving as a power-law \cite{pareto}, a result that
so far has not been disputed by different studies made across a wide
variety of samples obtained at different times for different
populations and in different countries or groups of countries
\cite[and references therein]{k80,nm09,fnm10,dy01, ferrero2,s05,yr09,
crbh08,by10, nm13,cyc05,c5}. Nevertheless, it has also been known for
quite some time that for the vast majority of the population, that
is, for those who do not belong to the very rich, this \textit{Pareto
power-law} does not hold.

Several studies can be found in the literature, especially in the
recent econophysical literature, with proposals regarding the behavior
of the income distribution of the whole population. Most of these
studies leave the Pareto power-law as the standard way of describing
the income data segment formed by the very rich, but model the income
data segment formed by the less rich by means of functions like the
exponential, the log-normal, the gamma function, the Gompertz curve,
as well as other functions \cite{dy01,cyc05,s05,nm09, fnm10,c5}.
Such approaches have been successful in terms of describing the
entire data range, but on the negative side they require fitting
the whole distribution with at least three parameters. In some
cases the number of parameters can go as high as five. In addition,
by dividing the income distribution in two segments one is in fact
assuming that societies are fundamentally divided in two very
distinct income classes, one formed by the rich, encompassing about
1\% of the population whose income is fitted by the Pareto power-law,
and another formed by the vast majority of people, the remaining
99\%, whose income is distributed according to other functions such
as the ones mentioned above. This methodology raises the question
of whether or not such a class division is really an intrinsic
feature of societies or just a result of fitting choices.  

A different approach for describing the income distribution was
made by Borges and Ferrero, who fitted income data
using the Tsallis functions instead of the ones mentioned above
and described the personal income distribution in terms of Tsallis'
$q$ parameter. Borges \cite{borges} fitted the income data to two
power-law regimes in two slopes, where one $q$ parameter controls
the slope of the first, intermediate, power-law regime and the
second $q$ parameter describes the tail of the distribution. In this
way he was able to describe almost the whole spectrum of the
\textit{county} distribution of the USA from 1970 to 2000, Brazil
from 1970 to 1996, Germany from 1992 to 1998 and United Kingdom from
1993 to 1998, concluding that in the case of the USA and Brazil an
increase in $q$ along the time indicates increasing inequality,
since greater values of $q$ imply greater probability of finding
counties much richer than others.

Ferrero's \cite{ferrero1,ferrero3} use of the Tsallis functions
to the income distribution problem showed to be possible to employ 
only one $q$ parameter for the whole income distribution range 
of a \textit{country}. He fitted income data from the UK, Japan, New
Zealand and USA, although the samples were limited to specific years
only: 1996 for New Zealand, 1998 for UK and Japan, and 2001 for the
USA. So, differently from Ref.\ \cite{borges}, his analysis did not
provide indications about the time evolution of the $q$ parameter,
although he concluded that for the first three countries $q$ is
close to 1.1 whereas it produces $q=1.29$ for the USA, a result which
seems to support Borges' conclusion that $q$ grows in parallel with
inequality, since among these countries the USA has the highest Gini
coefficient.

In this paper we deal with this old problem from the perspective of
Tsallis functions in an approach that combines and expands the analysis
of both authors above and reinforces them on the empirical side. We
propose using Tsallis statistics to represent the income distribution
of the \textit{entire income data range}, from the very poor to the super
rich, by a single function, that is, without assuming a class division.
We applied the Tsallis $q$-logarithm to the entire income data of the
Brazilian individual income distribution yearly samples from 1978 to
2014 using the same data reducing techniques previously applied in other
studies made with the personal income data of Brazil \cite{nm09,fnm10,nm13}.
This allowed us to study the time evolution of a \textit{single} $q$
parameter along a time span of almost four decades for the entire income
distribution of a whole \textit{country}, providing then new evidence of
Tsallis functions' ability to adequately represent personal income of a
whole country.

Our results show that Brazil's complementary cumulative
income distribution can be linearized by the $q$-logarithm and fitted by
using only two parameters. For Brazil $q$ ranges from 1.19 to 1.54 and
fluctuates with a period of approximately 3.5 years. Both fitted
parameters also present a cycling behavior in terms of one another
similar to the cycles obtained in Ref.\ \cite{nm13} through a
substantially different analysis where the method of describing the
income data range with two functions was applied. In addition, we noted
a second order effect, not previously reported in the study of any other
income samples, comprised of a periodic oscillation along the fitted
straight line whose amplitude grows with increasing income values.
Although such an effect can indeed be noted after a careful observation
of other income distribution studies made in different samples of different
countries using the method of dividing the income data in two domains
\cite{dy01,s05,nm09,by10}, it seems that it has not been previously
reported because this effect only becomes clearly visible when one fits
the entire income distribution range and reaches data values belonging
to the very rich. So, dividing the data in two functionally distinct
domains seems to obscure this periodic oscillation.

Considering that the entire income distribution range can be fitted
by just one function using only two parameters, a well-defined
two-classes-base income structure implicitly assumed when the income
range is described by two distinct functions may be open to questioning.
The point here is that such income-class division could possibly be
only a result of fitting choices and not of an intrinsic property of
societies. Although the Tsallis distribution is known to become a pure
power-law for large values of its independent variable $x$, and
exponential when $x$ tends to zero, this is not the same as assuming
from the start a two-classes approach to the income distribution
problem because the Tsallis distribution will only have power-law and
exponential like behaviors as limiting cases. Thus, a possible complex
behavior at the intermediate level might not be described by neither
of these functions. Hence, the Tsallis distribution does not necessarily
imply in two very distinct classes based on well-defined income domain
ranges, but possibly having an intermediate income range of unknown
size which might behave as neither of them. Although there may be
sociological evidences for this two-classes approach, from an
econophysical viewpoint, that is, from a modeling perspective, this is
an \textit{a priori} division because there is not yet a clear
\textit{dynamical} justification based on any known econophysical
\textit{mechanism} for doing that. So, from a dynamical viewpoint such
a class-based analytical approach might be a result of a purely fitting
procedure. It seems that this situation can only be clarified once we
have a full dynamical theory for the income distribution, theory which
is still lacking.

This paper is organized as follows. Sect.\ 2 presents the Tsallis
functions and some results based on them required in our analysis.
Sect.\ 3 presents and discusses our fitting results with several
graphs. Sect.\ 4 presents our conclusions.

\section{Tsallis functions}

The Tsallis statistics is based on the $q$-logarithm and $q$-exponential
functions, defined as follows \cite{tsallis1,tsallis2},
\be
\ln_q x \equiv \frac{x^{(1-q)}-1}{1-q},
\lb{qlog}
\ee
\be
{{\mathrm{e}}_q}^x \equiv {\left[1+(1-q)x \right]}^{1/(1-q)}.
\lb{qexp}
\ee
For $q=1$ both functions become the usual logarithm and exponential,
that is, ${\mathrm{e}_1}^x={\mathrm{e}}^x$ and $\ln_1 x=\ln x$. Hence,
Tsallis $q$-functions are in fact the usual exponential and logarithmic
functions deformed in such a way as to be useful in Tsallis' theory of
nonextensive statistical mechanics \cite{tsallis2}. Nevertheless, they
are not the only way of deforming these two common functions in order
to suit specific applications, which include, among others, the personal
income distribution. Another way of doing this is by employing the
$\kappa$-generalized exponential, which was in fact advanced as a single
function capable of fitting the whole income data range as, similarly to
the Tsallis $q$-functions, it has the exponential and power-law as a
limiting cases \cite{cgk07}. Since it has been extensively studied
elsewhere in the context of income and wealth distributions \cite{cmgk08,
cgk09,cgk12} we shall not deal with it in this paper.

From their very definitions, it is clear that,
\be
{{\mathrm{e}}_q}^{(\, \ln_q x)}= \ln_q ({\mathrm{e}_q}^x \,)=x.
\lb{recip}
\ee
In addition, $\ln_q 1=0$ for any $q$. So, if there exists a value
$x_0$ such that $x/x_0=1$ then $\ln_q (x/x_0)=0$. Two other
properties of the $q$-exponential useful in the present context
are written below \cite{yamano},
\be
{\left[ {\mathrm{e}_q}^{f(x)} \right] }^{\up{a}}=
{\mathrm{e}}^{\up{a f(x)}}_{\dn{\dn{{1-(1-q)/a}}}} \, ,
\lb{qexpa}
\ee
\be
\frac{\dd}{\dd x} \left[ {\mathrm{e}_q}^{f(x)} \right]=
\dfrac{{\left[ {\mathrm{e}_q}^{f(x)} \right]
}^{q}}{f {\displaystyle \,'}(x)}.
\lb{dfeqfx}
\ee

\section{Income distribution}

Let $\mathcal{F}(x)$ be the \textit{cumulative distribution function}
(CDF) of individual income, which gives the probability that an
individual receives an income \textit{less than or equal to} $x$.
Hence, the \textit{complementary cumulative distribution function}
(CCDF) of individual income $F(x)$ gives the probability that an
individual receives an income \textit{equal to or greater than}
$x$. From this it is clear that $\mathcal{F}(x)+F(x)=100$, where
the maximum probability is normalized as 100\% instead of the
usual unity value. These functions have the following approximate
boundary conditions \cite{nm09,fnm10}, $\mathcal{F}(0)= {F}(\infty)
\cong 0$ and $\mathcal{F}(\infty)= {F}(0) \cong 100$. Besides,
$\dd\mathcal{F}(x)/\dd x = - \dd F(x)/\dd x=f(x)$ and
$\int_0^\infty f(x)\: \dd x=100$, where $f(x)$ is the
\textit{probability density function} (PDF).

The connection of the Tsallis functions with income distribution
comes from the realization that when $F(x)$ is plotted in a
log-log scale, its functional behavior of decreasing values as 
the income $x$ increases is very similar to the behavior of
${\mathrm{e}_q}^{-x}$ for $q>1$ also plotted in a log-log scale
(see Fig.\ 3.4 of Ref.\ \cite{tsallis2}). In addition, the Tsallis
functions behave as a power-law for high income values, that is, at
the tail of the distribution \cite{ferrero1}. So, this suggests that
we can describe the income distribution by means of the following
expression,
\be
F(x)=A \, {\mathrm{e}_q}^{-Bx},
\lb{Fexpq}
\ee
where $A$ and $B$ are positive parameters. Considering the boundary
condition $F(0)=100$ one can straightforwardly conclude that $A=100$.
Hence, substituting this result into the expression above and taking
the $q$-logarithm we obtain,
\be
\ln_q \left[ \dfrac{F(x)}{100} \right]=-B \, x.
\lb{Flnq}
\ee
Considering Eqs.\ (\ref{qexpa}) and (\ref{dfeqfx}), the corresponding
PDF yields,
\be
f(x)=\dfrac{100}{B}{\left({\mathrm{e}_q}^{-Bx}\right)}^q=
\dfrac{100}{B} \left( \mathrm{e}_{\dn{2-1/q}}^{\up{-qBx}} \right).
\lb{pdf}
\ee
Eq.\ (\ref{Flnq}) clearly goes through the coordinate's origin if
we remember the properties of the $q$-logarithm outlined above. So,
the data fitting problem is then reduced to finding only two
parameters, $q$ and $B$. This has been done for all Brazilian
samples constituted by a time span of almost fours decades, as
explained in what follows.

To calculate the optimal $q$ parameter, it is important to bear
in mind that every yearly sample from 1978 to 2014 produced an
empirical set of $n$ observed income values $x_i, (i=1,\ldots,n)$
and their correspondent CCDF $F_i=F(x_i)$. To find the optimal $q$
parameter for a specific yearly dataset $\{F_i,x_i\}$, we assumed
that in each annual sample the optimal $q$ lies in the interval
$[-10,10]$ and then ranged it with steps of $\Delta q=0.003$.
Hence, we generated another set of $m$ values $q_j$ $(j=1\ldots,m)$,
fitted a straight line to the specific dataset under study for each
$q_j$ to obtain the correspondent fitted parameter $B_j$. In this way
each year produced another set of $m$ quantities $\{q_j,B_j\}$. For 
each pair of values $q_j,B_j$ we defined the following residue,
\be
R_j (q_j) = \sum_{i=1}^n \Bigg[ \ln_{q_j} \Bigg( \frac{F_i}{100}
\Bigg) + B_j \, x_i \Bigg].
\lb{residuo}
\ee
In view of Eq.\ (\ref{Flnq}) the ideal $q$ implies in an ideal $B$ and
in turn they both produce zero residue. So the pair of parameters that
produces the minimum residue value $R_j$ are the optimal $q_j$ and
$B_j$ of the sample. Finding the optimal pair of parameters is an
extremum problem, which means that the optimal pair $q_j$, $B_j$
produces the minimum $R_j$ and that corresponds to the maximum value of
the second derivative ${\dd}^2 R_j /{\dd {q_j}}^2$. By following this
procedure computationally we were able to find the best fit for both
$q$ and $B$ in each year for all our samples. The step interval
$\Delta q$ gave us a rough indication of the uncertainty in the optimal
$q$ and the error in $B$ had been previously obtained from the usual
linear fit.

Figs.\ \ref{mosaic1}--\ref{mosaic6} show graphs of Brazil's 
CCDF linearized, straight line fitted according to Eq.\ (\ref{Flnq})
and plotted against the income $x$. The fitted parameters are
summarized in Table \ref{tab1}. It is clear that the $q$-functions
are very successful in fitting the whole distribution, and it is
also clear that there is another, second order, effect consisting
of a periodic oscillation of the data along the fitted straight line.
This effect has not been reported before in income distribution
studies, perhaps because it only becomes clearly visible when one
fits the whole data range as the amplitude of the oscillation grows
with the income. It becomes more prominent at the tail of the
distribution, that is, where it behaves more as a power-law.
Actually, this oscillatory behavior can indeed be observed once one
takes a careful look at previous studies of income distribution made
with different samples at different time periods and using
different methodologies, since even low amplitude oscillations are
also present at very low income data. This can be verified,
\textit{e.g.}, in Figs.\ 8 and 9 of Ref.\ \cite{nm09}, at insets of
Figs.\ 1 and 2 of Ref.\ \cite{dy01}, and in Fig.\ 27 of Ref.\
\cite{s05}, although none of these studies have actually reported
the presence of this periodic oscillation in their data.

Fig.\ \ref{q-B-year} shows the time evolution of both $q$ and $B$
along the time span of this study. There is a clear periodicity in
their temporal variation, with maxima appearing from 2 to 5 years,
actually at about 3.5 years on average. Such periodicity in both
parameters is not a novelty as far as Brazilian income data are
concerned, since it has also appeared in other treatments even when
different methodologies were applied \cite{nm09,fnm10, nm13}. Hence,
it is conceivable that those different methodologies could be unified
with the present one by means of possible relationships of the
parameters used in those different studies.

Fig.\ \ref{B-q} shows both parameters plotted in terms of one another.
Although the points present some dispersion, there is a tendency for
$B$ to grow linearly with $q$, although this pattern is unrelated to
time. This tendency can be better seen by a straight line weighted fit
to the data, shown as a dashed line in the figure, which indeed
indicates a growing pattern. However, a time related pattern does appear
in Fig.\ \ref{q-B-ciclos}, where the data were divided in three time
intervals, 1978--1989, 1990--2001 and 2002--2014, so that this
pattern becomes more clearly visible. By following the points
chronologically along the dashed lines a cycle appears, mostly clockwise,
but with a few anti-clockwise turns. What is striking about these plots
is their similarities with the ones discussed by Moura Jr.\ and Ribeiro
\cite{nm13} where a clockwise cycle is also present in their study,
although this was a result of an analysis employing an entirely
different methodology (see Fig.\ 3 of Ref.\ \cite{nm13}). Fig.\
\ref{B-q-year-3D.eps} shows the same results but in three dimensions
where the fitted parameters evolve along a helix like line. 

These results taken together do seem to indicate a nontrivial dynamics
in the income distribution evolution whose origins are still basically
unknown, although Ref.\ \cite{nm13} provided some possible indications
of its origin. More studies are necessary in order to better understand
the dynamical significance of the periodic oscillation along the fitted
$q$-logarithm (Figs.\ \ref{mosaic1}--\ref{mosaic6}), their
periodicity (Fig.\ \ref{q-B-year}), time unrelated growth pattern
(Fig.\ \ref{B-q}), temporal cycling features (Fig.\ \ref{q-B-ciclos})
and helical like evolution (Fig.\ \ref{B-q-year-3D.eps}). Nevertheless,
since some of these features can be observed in different samples of
different countries whose data were fitted by different functions by
means of different methodologies, this indicates that these effects are
real and deserve further investigation.

\begin{table*}[t]
\caption{Fitted values of the parameters of Eq.\ (\ref{Flnq}) for 
         Brazilian income data from 1978 to 2014.\label{tab1}}
\begin{center}
\small
\begin{tabular}{ccccccccccc}
\hline\noalign{\smallskip}
year & $ q (\pm0.003) $ & $B$ \\ 
\noalign{\smallskip}\hline\noalign{\smallskip}
1978 & $1.397$ & $0.484\pm0.065$ \\
1979 & $1.322$ & $0.421\pm0.067$ \\
1981 & $1.235$ & $0.994\pm0.034$ \\
1982 & $1.418$ & $2.462\pm0.088$ \\
1983 & $1.238$ & $0.838\pm0.050$ \\
1984 & $1.253$ & $1.124\pm0.034$ \\
1985 & $1.241$ & $0.779\pm0.039$ \\
1986 & $1.382$ & $1.234\pm0.112$ \\
1987 & $1.424$ & $2.133\pm0.095$ \\
1988 & $1.247$ & $0.838\pm0.044$ \\
1989 & $1.397$ & $1.342\pm0.058$ \\
1990 & $1.490$ & $3.312\pm0.155$ \\
1992 & $1.415$ & $1.737\pm0.113$ \\
1993 & $1.397$ & $1.564\pm0.070$ \\
1995 & $1.244$ & $0.846\pm0.039$ \\
1996 & $1.238$ & $0.799\pm0.045$ \\
1997 & $1.361$ & $1.627\pm0.075$ \\
1998 & $1.328$ & $1.369\pm0.033$ \\
1999 & $1.301$ & $1.250\pm0.043$ \\
2001 & $1.187$ & $0.549\pm0.055$ \\
2002 & $1.352$ & $1.676\pm0.059$ \\
2003 & $1.292$ & $1.229\pm0.038$ \\
2004 & $1.292$ & $1.216\pm0.045$ \\
2005 & $1.382$ & $1.910\pm0.108$ \\
2006 & $1.229$ & $1.032\pm0.047$ \\
2007 & $1.349$ & $1.414\pm0.077$ \\
2008 & $1.313$ & $1.298\pm0.060$ \\
2009 & $1.511$ & $4.551\pm0.271$ \\
2011 & $1.379$ & $2.033\pm0.081$ \\
2012 & $1.538$ & $3.089\pm0.335$ \\
2013 & $1.265$ & $0.998\pm0.068$ \\
2014 & $1.265$ & $0.958\pm0.067$ \\
\noalign{\smallskip}\hline
\end{tabular}
\end{center}
\end{table*}

\section{Conclusions}

In this paper we have used the Tsallis functions $q$-exponential and
$q$-logarithm to describe the personal income data of Brazil. Yearly
samples from 1978 to 2014 were linearized using the $q$-logarithm and
fitted to a straight line, providing then a single function
representation of the whole distribution using only two parameters.
A second order effect not previously reported was clearly noticed in
the form of a periodic oscillation of the data around the fitted
straight line, whose amplitude steadily grows with increasing income
values to finally become clearly visible at the tail of the
distribution. In addition, the fitted parameters tend to grow in terms
of one another, growth which is unrelated to time, but they do present
a time related feature since they cycle chronologically in terms of
one another in a general clockwise pattern with a few anti-clockwise
turns.

As mentioned in Sect.\ \ref{intro}, the Tsallis functions have been
previously used by Borges \cite{borges} and Ferrero \cite{ferrero1,
ferrero3} to describe the personal income distribution, although the
former applied them only to the intermediate and tail portions of the
distribution of \textit{county} income data, whereas the latter was
applied to samples derived from the income of whole \textit{countries},
but limited to some specific years only. Here we applied the Tsallis
functions to the entire population of Brazil for a time span of almost
four decades and fitted the functions to the whole distribution,
without subdivisions. However, some of the conclusions reached here
are similar to the ones reached by these authors, particularly the
suitability of the Tsallis functions to adequately describe the income
distribution of a population and that an increase of the $q$ parameter
seems to imply greater inequality. Indeed, since Brazil has in general
higher values for both the Gini coefficient and the $q$ parameter as
compared to the ones produced by the countries studied by these authors,
USA, New Zealand, UK, Japan and Germany, these two coefficients seem to
behave similarly: higher values of $q$ appear to imply greater income
distribution inequality.

Notwithstanding, the observed oscillatory behavior in the linearized
complementary cumulative data distribution is a new effect which
ought to be considered in future studies of income distribution. It is
known that power-law like distributions, which behave as pure power
distribution for large values of the independent variable $x$ and
exponential for $x \rightarrow 0$, can be identified with the Tsallis
distribution as given by Eq.\ (\ref{Fexpq}). The nontrivial aspect of
this distribution is that in different parts of the space defined by
the variable $x$ one finds the dominance of different dynamical factors.
In addition, as discussed by Wilk and W{\l}odarczyk \cite{wilk}, there
are experimental results and empirical observations that can be
described by a Tsallis distribution and which exhibit log-periodic
oscillations, such as earthquakes \cite{huang} and stock markets near
financial crashes \cite{sornette,vande1,vande2,wosnitza} to name just two
of these observed structures \cite[see Ref.][and references therein]{wilk}.
The point is that such oscillating factors are visible in these processes,
but are somehow hidden in the distribution given by Eq.\ (\ref{Fexpq}).
When taken into account, such oscillations usually ``decorate'' the PDF
(\ref{pdf}) by multiplying it with some log-periodic oscillating factor.
What is interesting in the approach of Ref.\ \cite{wilk} is that such
oscillations are introduced into Tsallis distributions by allowing the
$q$ parameter to become complex. This happens at the cost of introducing
further parameters in the description, but that seems inevitable since
such weak, but persistent, oscillating structures in the data indicate
that the system under study has scale-invariant behavior and their presence
imply into the existence of important dynamical features hidden in the
fully scale-invariant description. The consequences for considering these
oscillating features into the income distribution problem are still
unknown, but if the work of Ref.\ \cite{wilk} could be taken as a
possible template on how to look at this problem they may imply in
important constraints on the underlying income distribution dynamics.
Moreover, considering that periodicities do not appear only along the
distribution, but are also present in the fitted parameters themselves
once they are related to one another, as shown by their chronological
cycling behavior along their time evolution, all these features taken
together clearly indicate the existence of a nontrivial income
distribution dynamics whose origins are unclear and, therefore,
deserve further investigation.

Finally, these results bring further questions about the traditional
way of representing personal income data by splitting them in two
segments, one for the very rich, described by the Pareto power-law,
and another for the rest of the population, described by other
functions. Such a segmentation forms the basis of the claims that
societies are fundamentally structured in a two-classes-income system.
The point is that if only one function is able to describe the whole
distribution, albeit this function tends to the exponential at low
income values and power-law at large ones, such a very well defined
class based structure might not be the single most essential feature
of societies, but might have its prominence as just a result of a
fitting methodology. Hence, there might be an intermediary income
range of unknown size whose dynamics may be crucial in the understanding
of income dynamics. The point here is that although there may be
sociological evidence for a two-classes approach, it might be argued
that societies usually have a third, intermediate, middle income
segment, known generically as ``middle class'', whose dynamics, that is,
whose income structure evolution, possibly oscillates between two
extremes and be responsible for such oscillatory behaviors. These
points may only be clarified once one has a full dynamical theory of
income distribution, theory which is still lacking.

\begin{figure}[ht]
\begin{center}$
\begin{array}{ccc}
\includegraphics[scale=0.4]{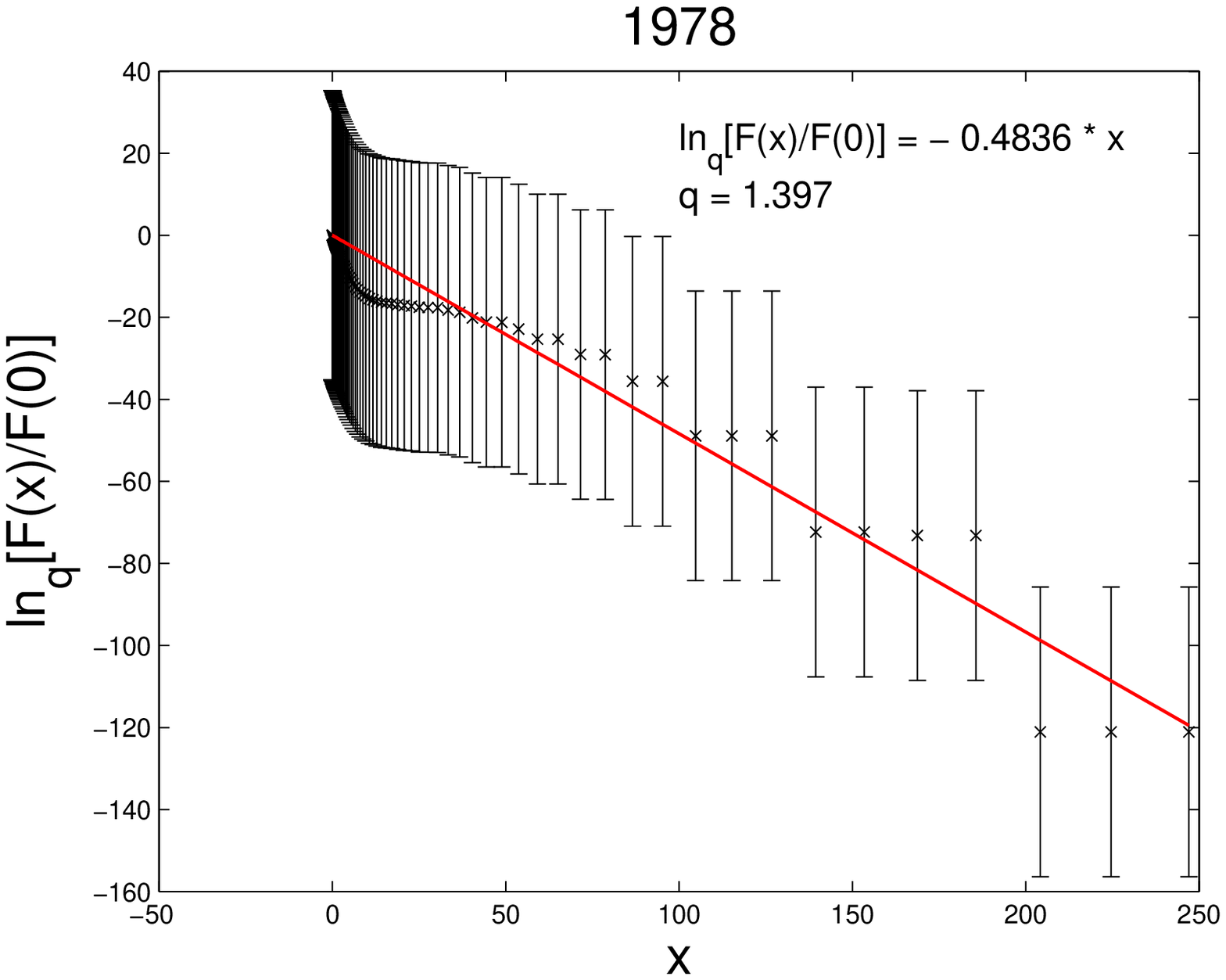} &
\includegraphics[scale=0.4]{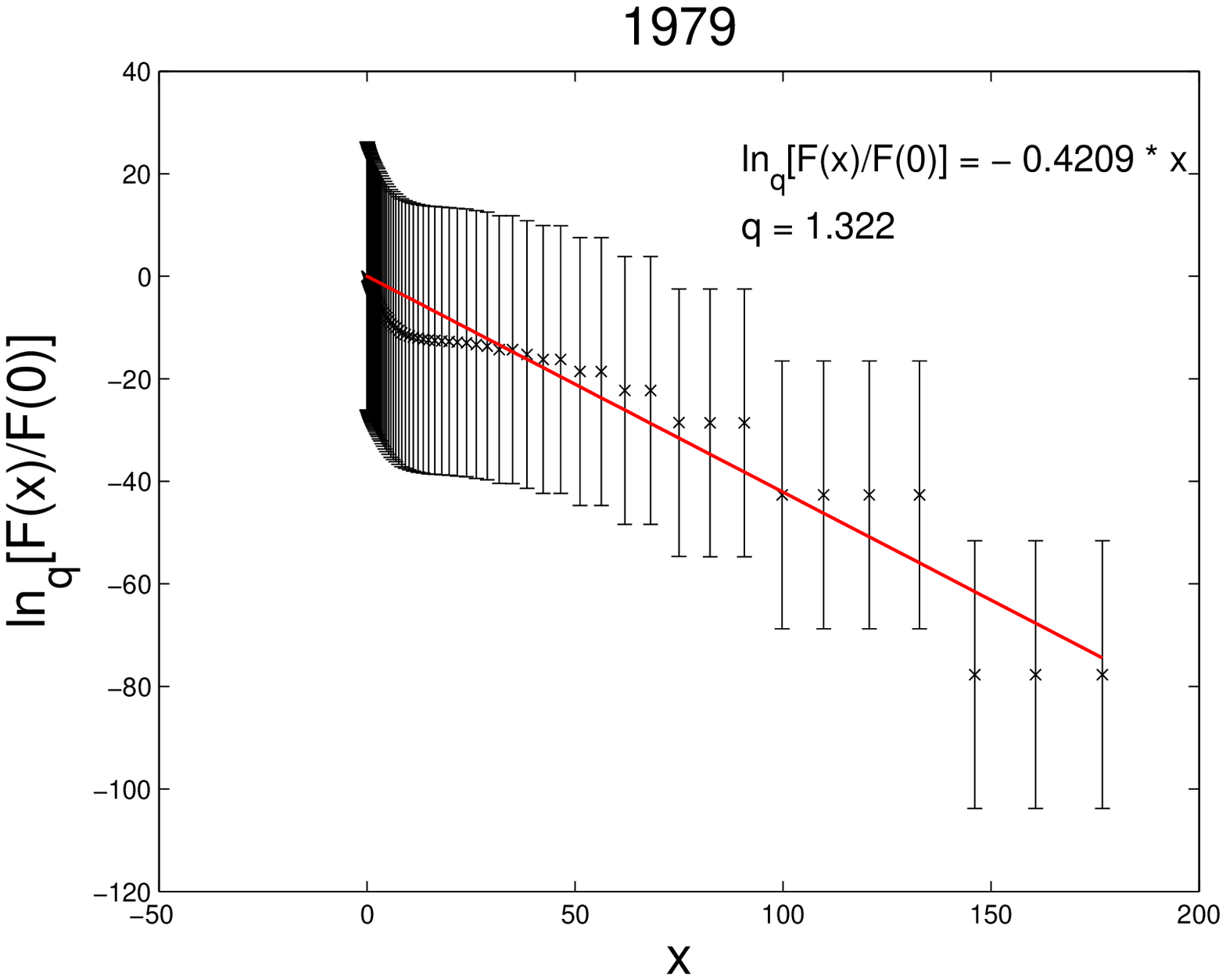} \\ 
\includegraphics[scale=0.4]{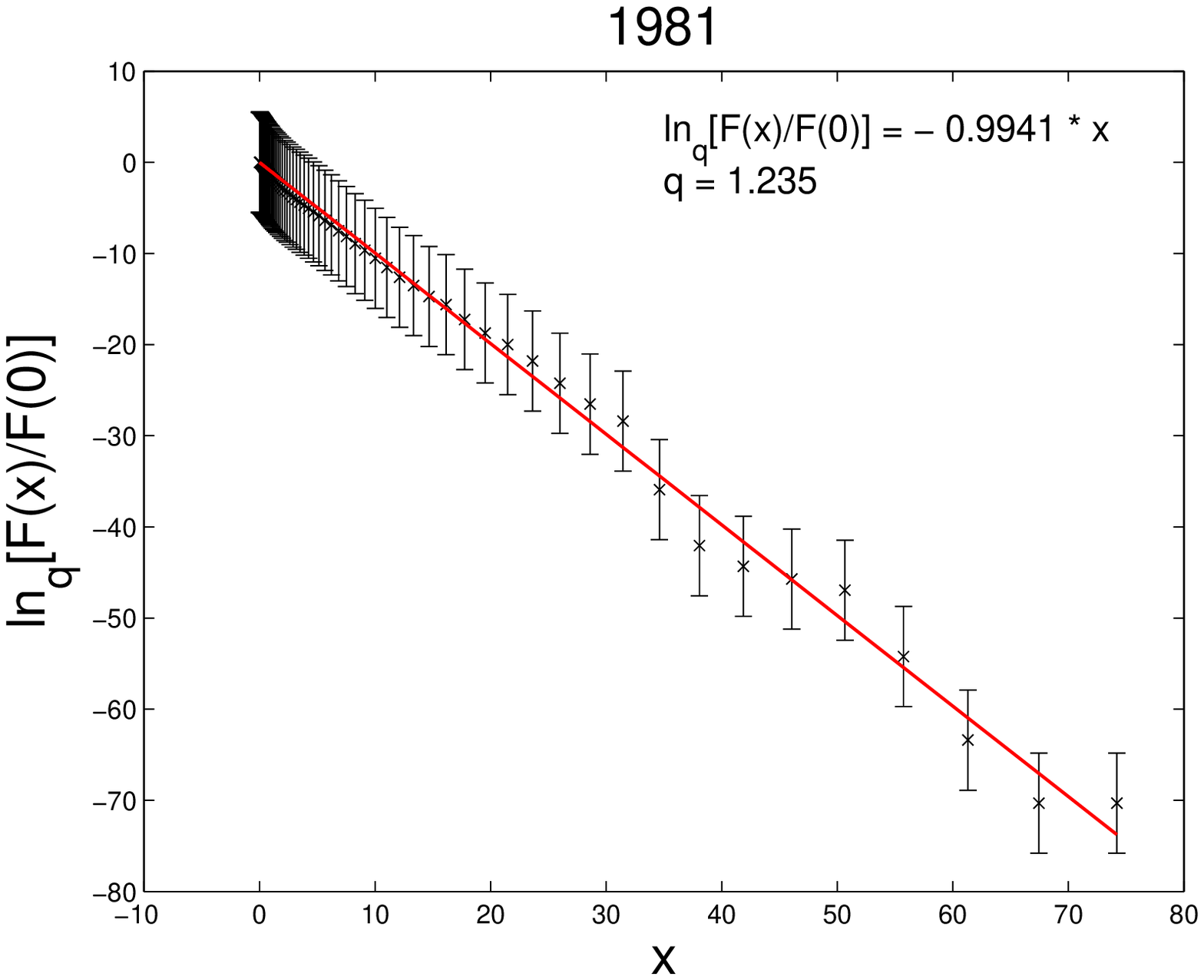} &
\includegraphics[scale=0.4]{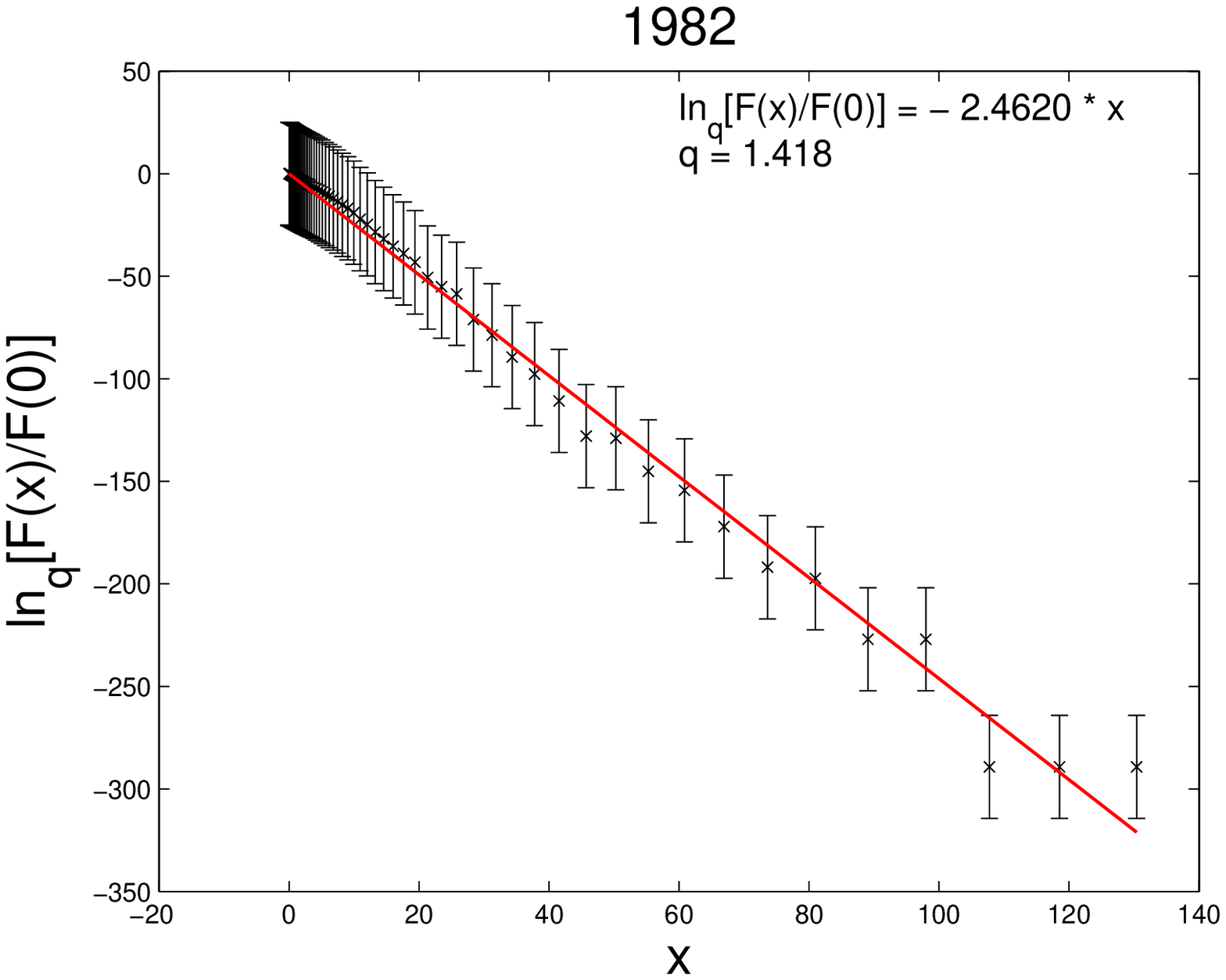} \\
\includegraphics[scale=0.4]{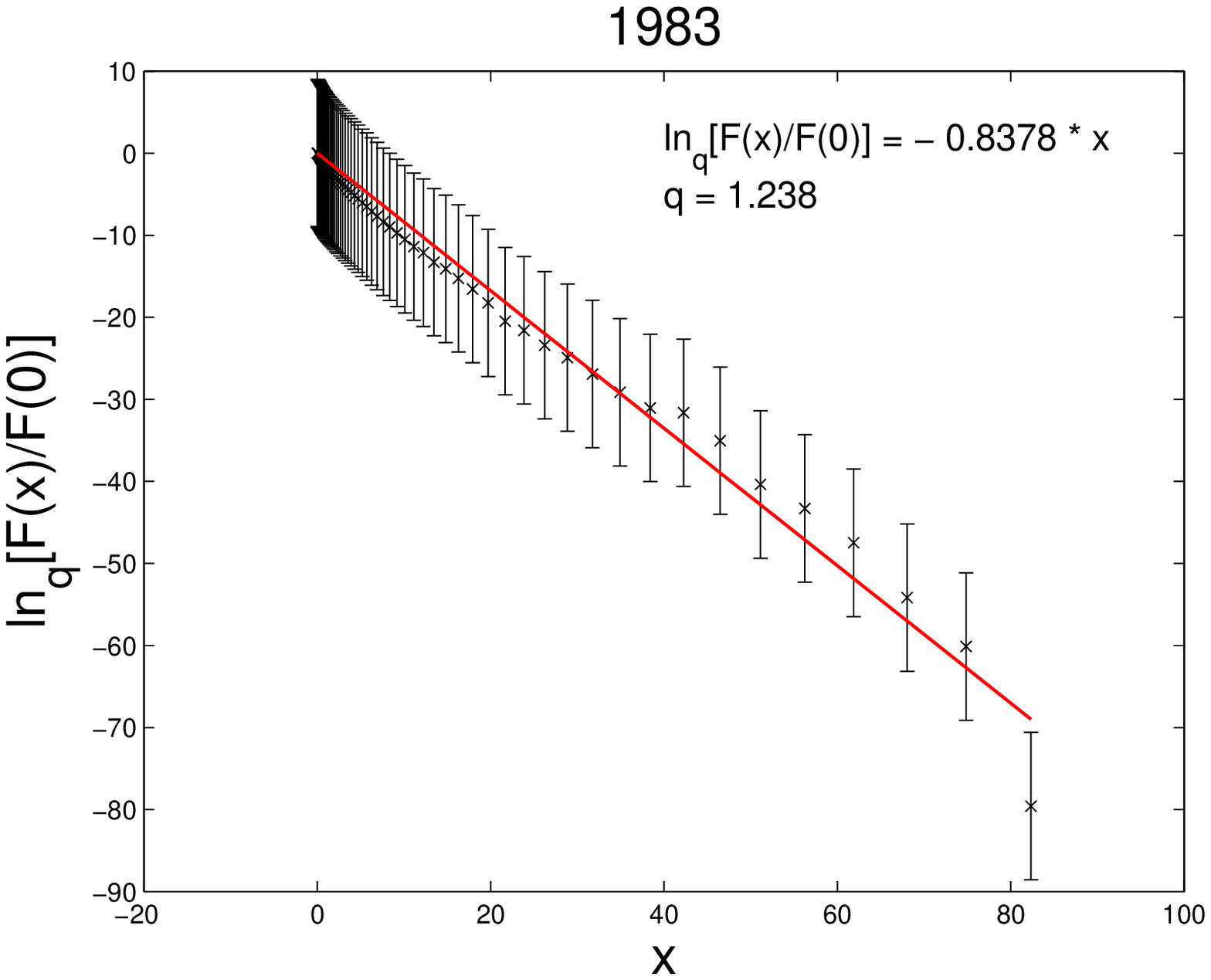} & 
\includegraphics[scale=0.4]{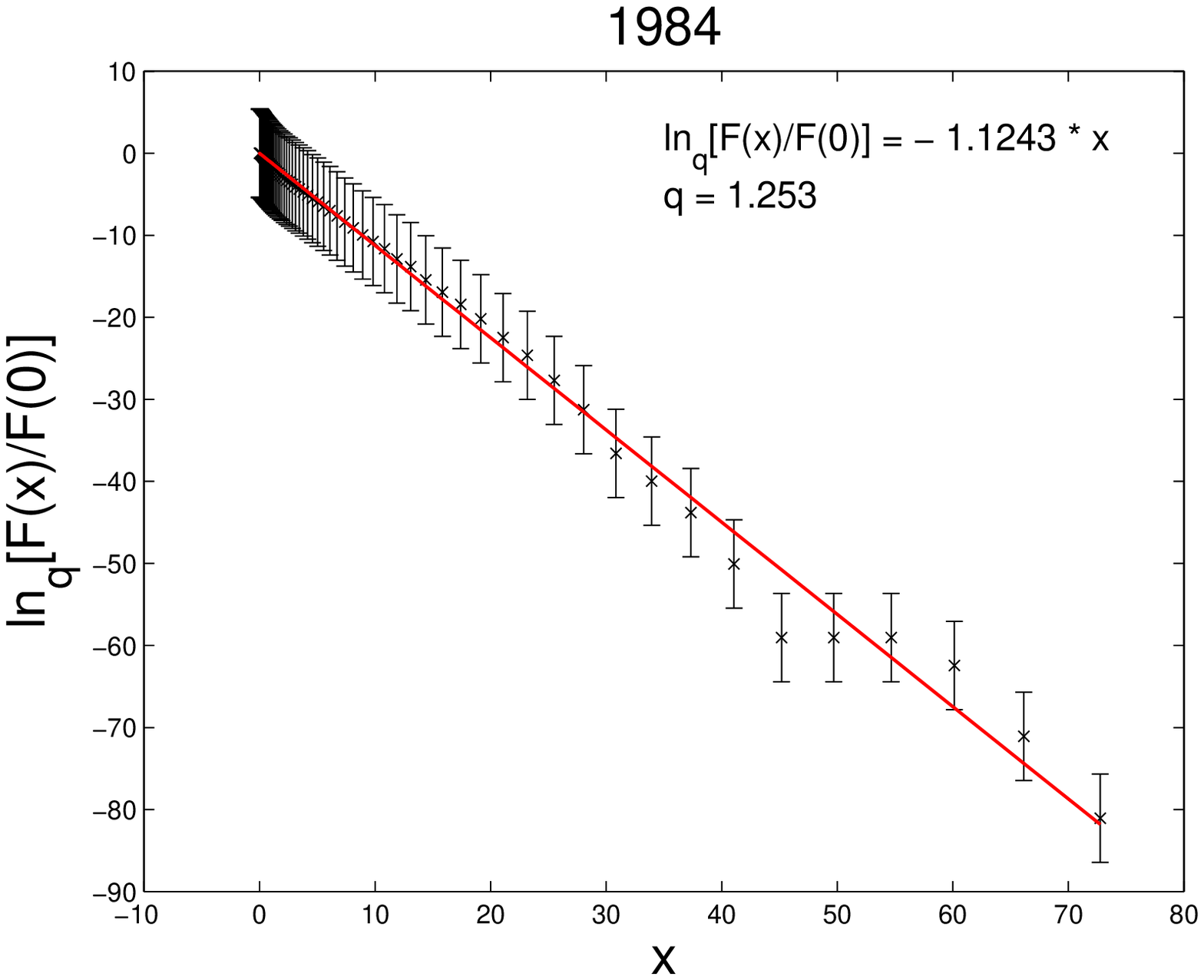} 
\end{array}$
\end{center}
\caption{Graphs of the linearized CCDF against the income $x$
according to Eq.\ (\ref{Flnq}) of the yearly samples of Brazilian
personal income data from 1978 to 1984. The income $x$ is given in
terms of the average income of the respective year, that is, a value
of, say, $x=10$ means 10 times the average income \cite[see Ref.][]{nm09}.
One can observe that the data oscillate around the fitted straight
line with an amplitude that steadily grows with increasing income
values.}
\lb{mosaic1}
\end{figure}

\begin{figure}[ht]
\begin{center}$
\begin{array}{ccc}
\includegraphics[scale=0.4]{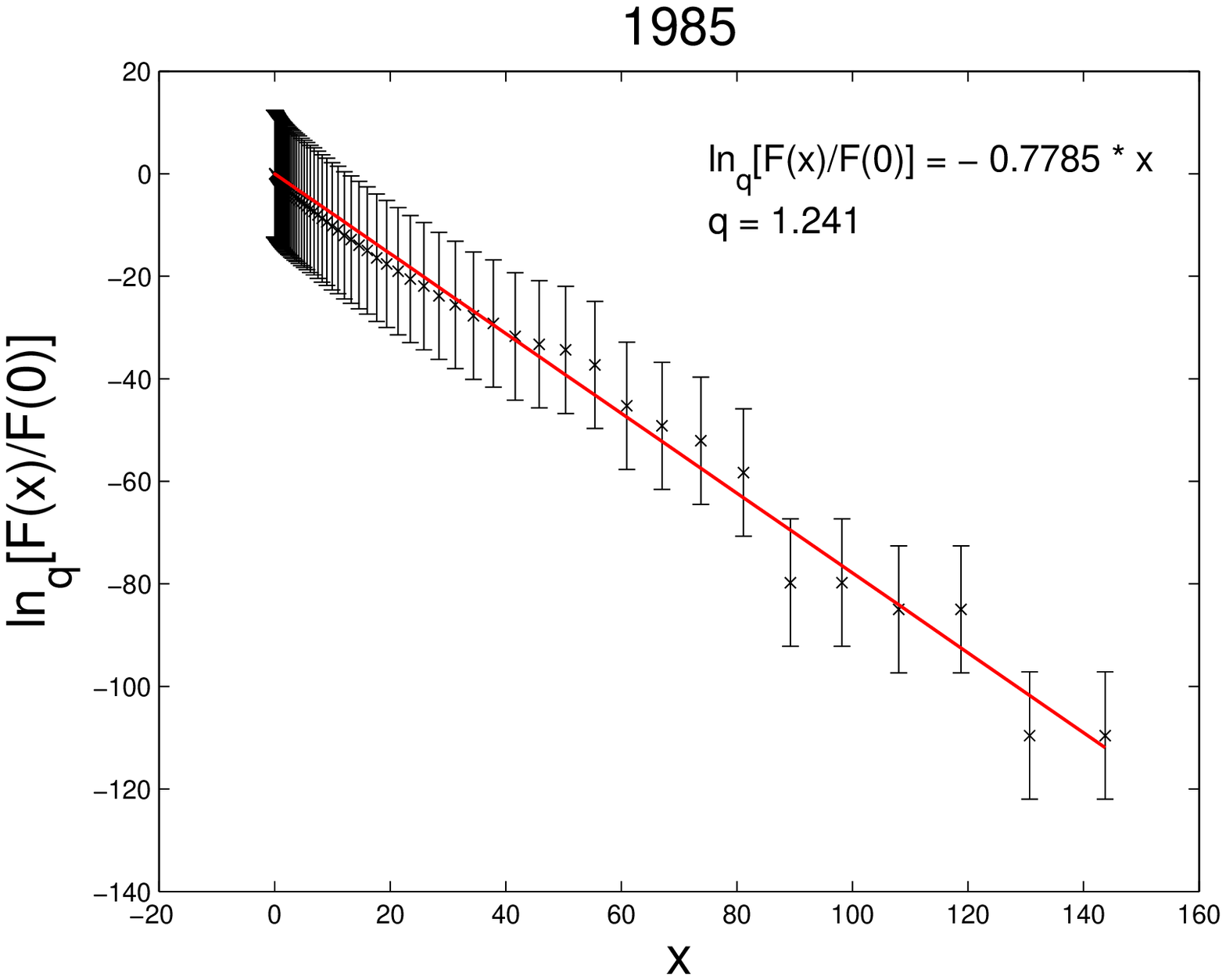} &
\includegraphics[scale=0.4]{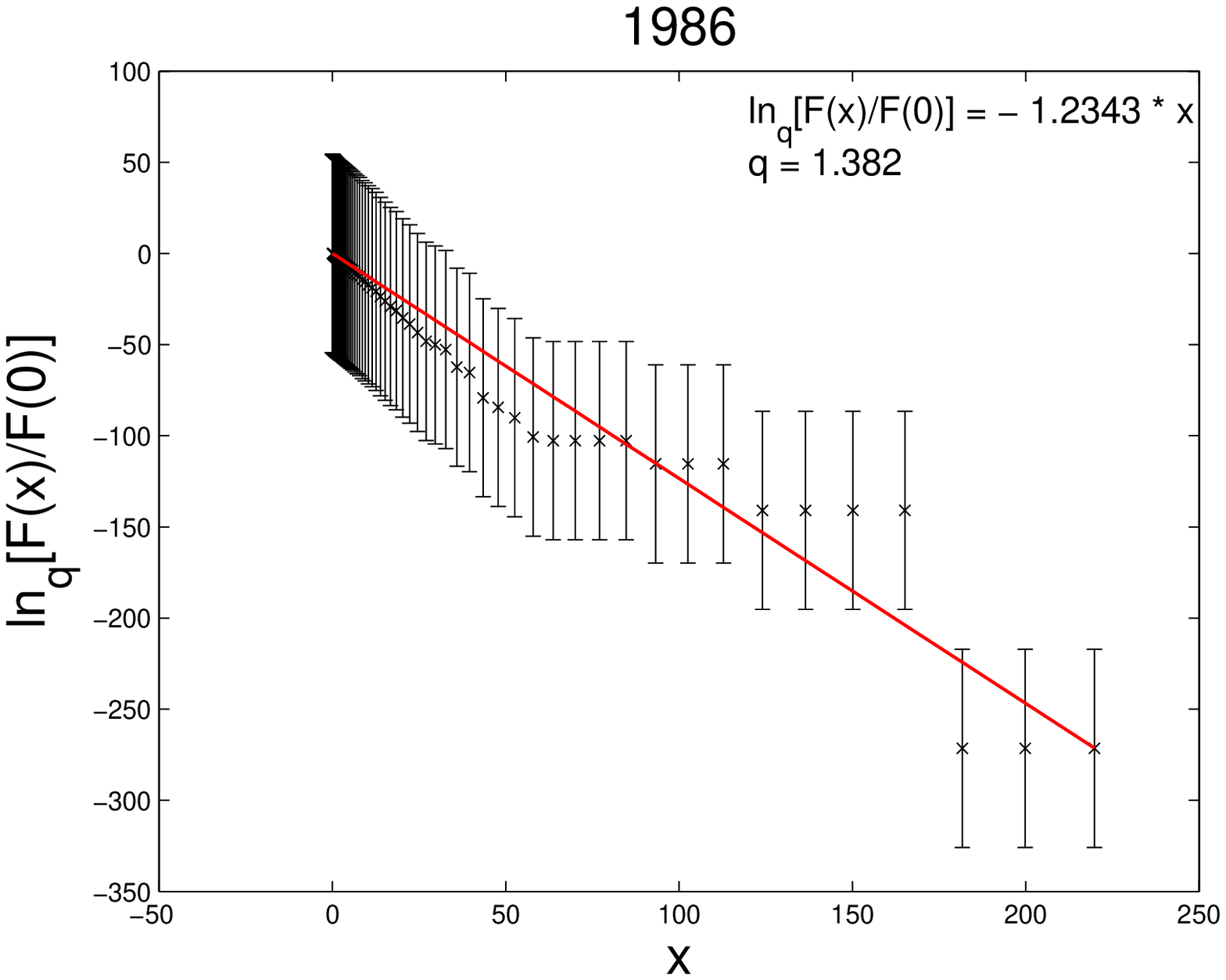} \\ 
\includegraphics[scale=0.4]{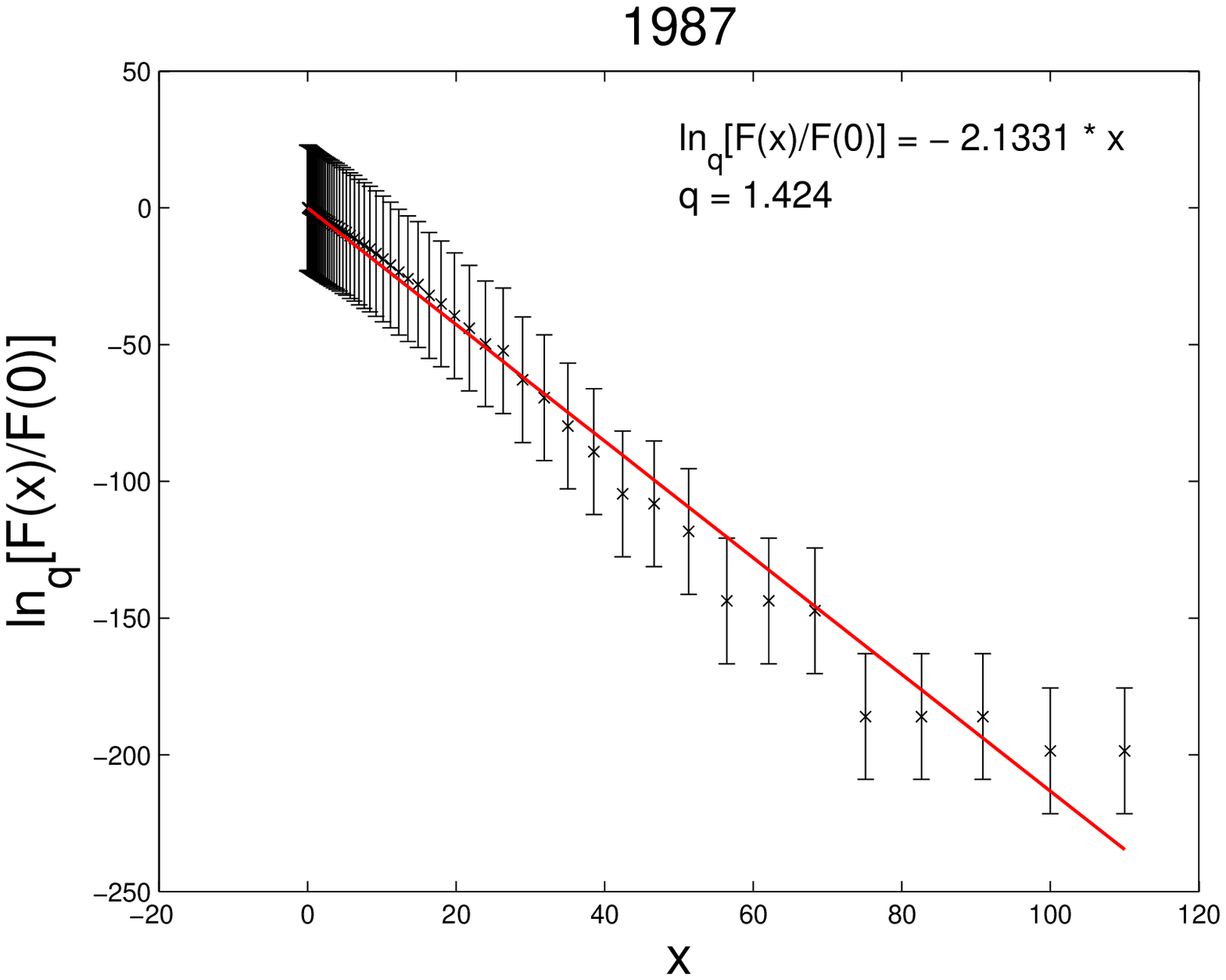} &
\includegraphics[scale=0.4]{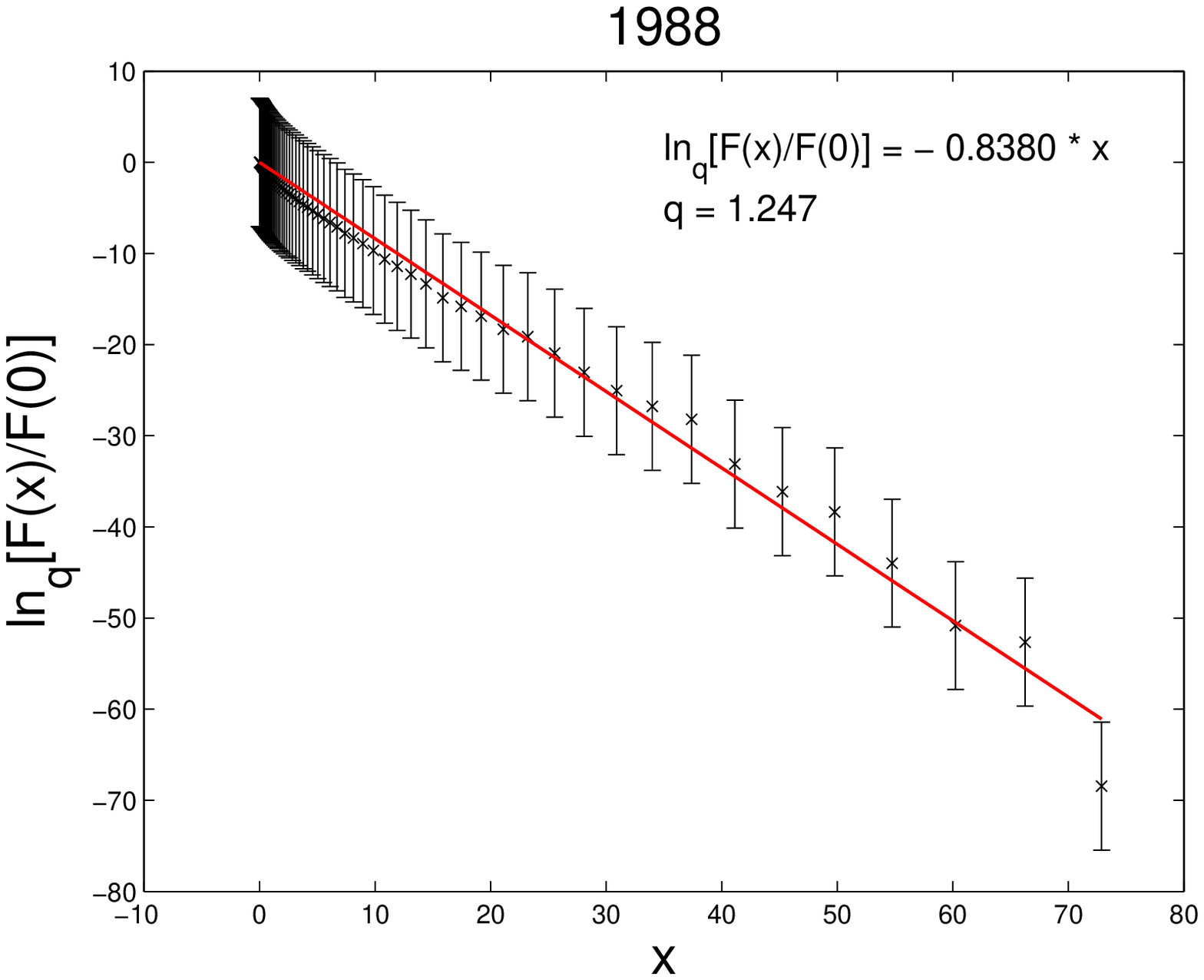} \\
\includegraphics[scale=0.4]{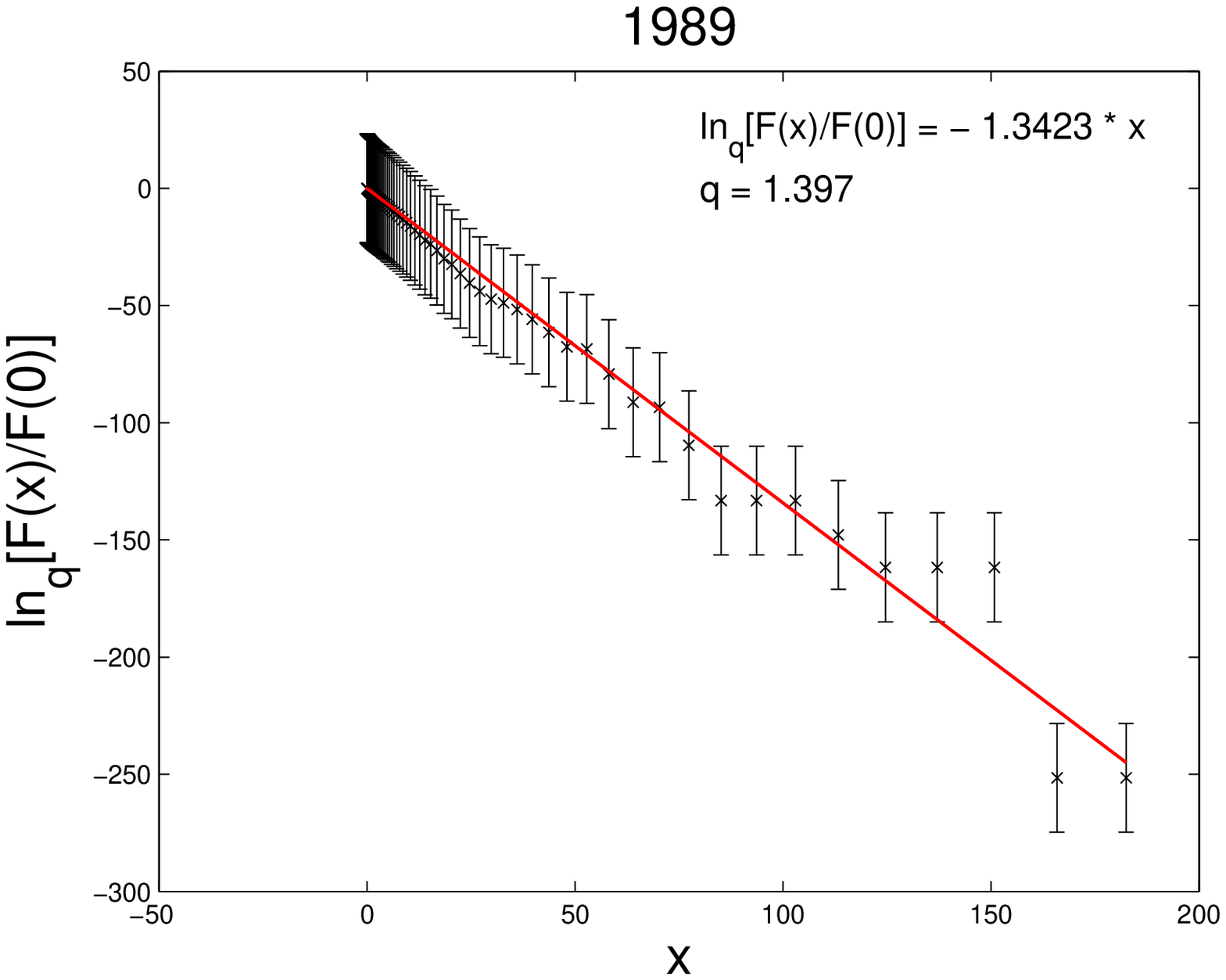} & 
\includegraphics[scale=0.4]{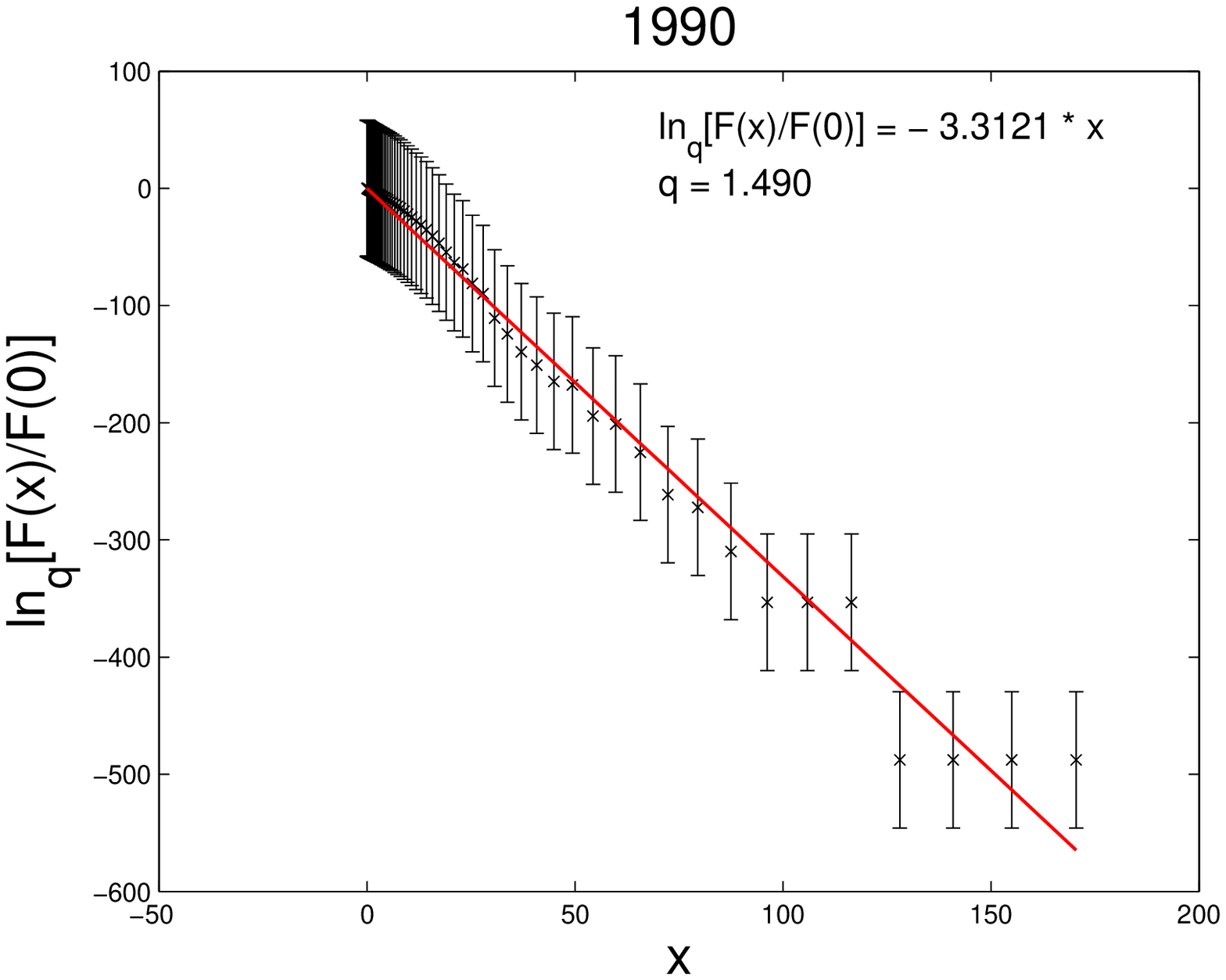} 
\end{array}$
\end{center}
\caption{Continuation of the previous graphs with data from 1985 to 1990.}
\lb{mosaic2}
\end{figure}

\begin{figure}[ht]
\begin{center}$
\begin{array}{ccc}
\includegraphics[scale=0.4]{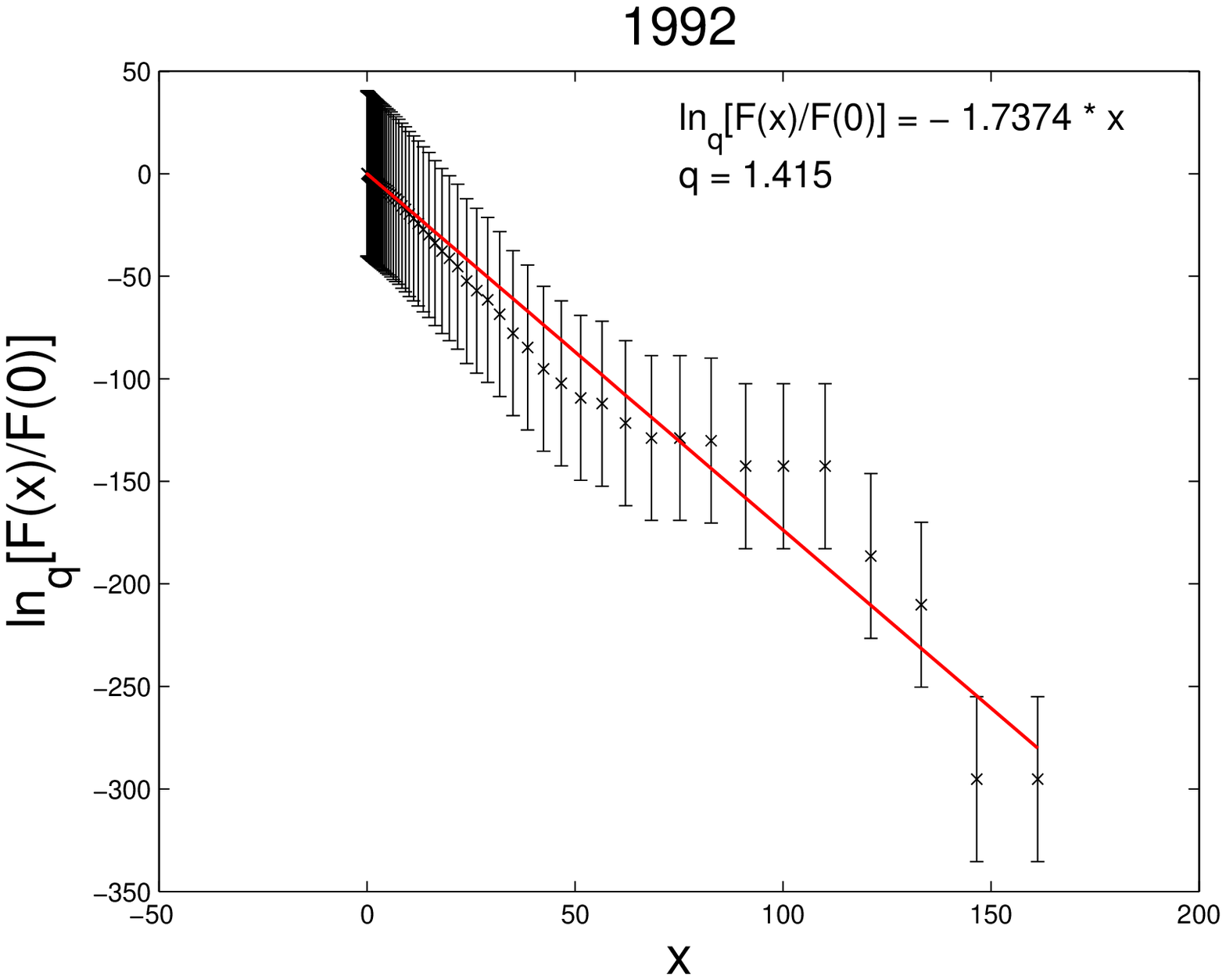} &
\includegraphics[scale=0.4]{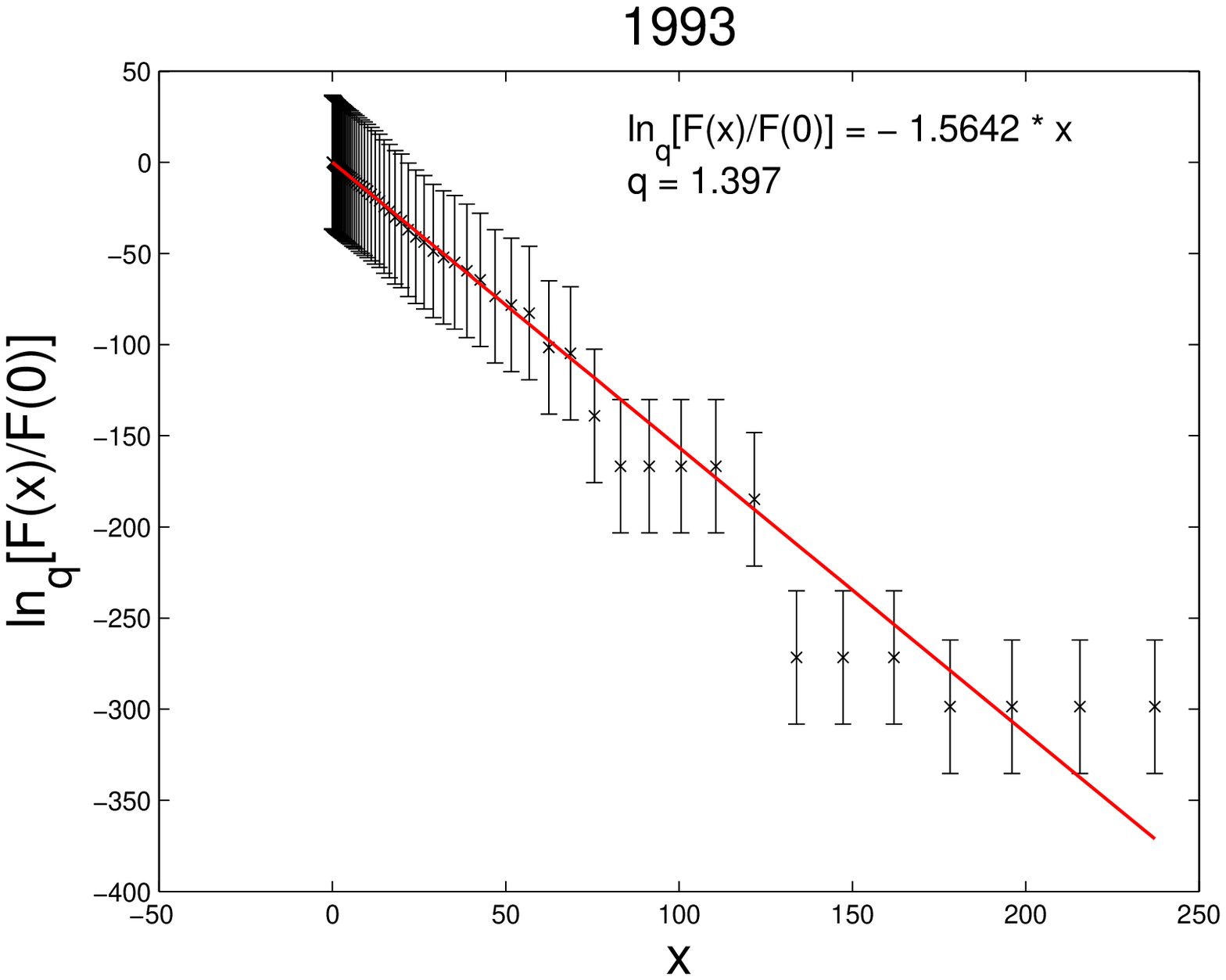} \\ 
\includegraphics[scale=0.4]{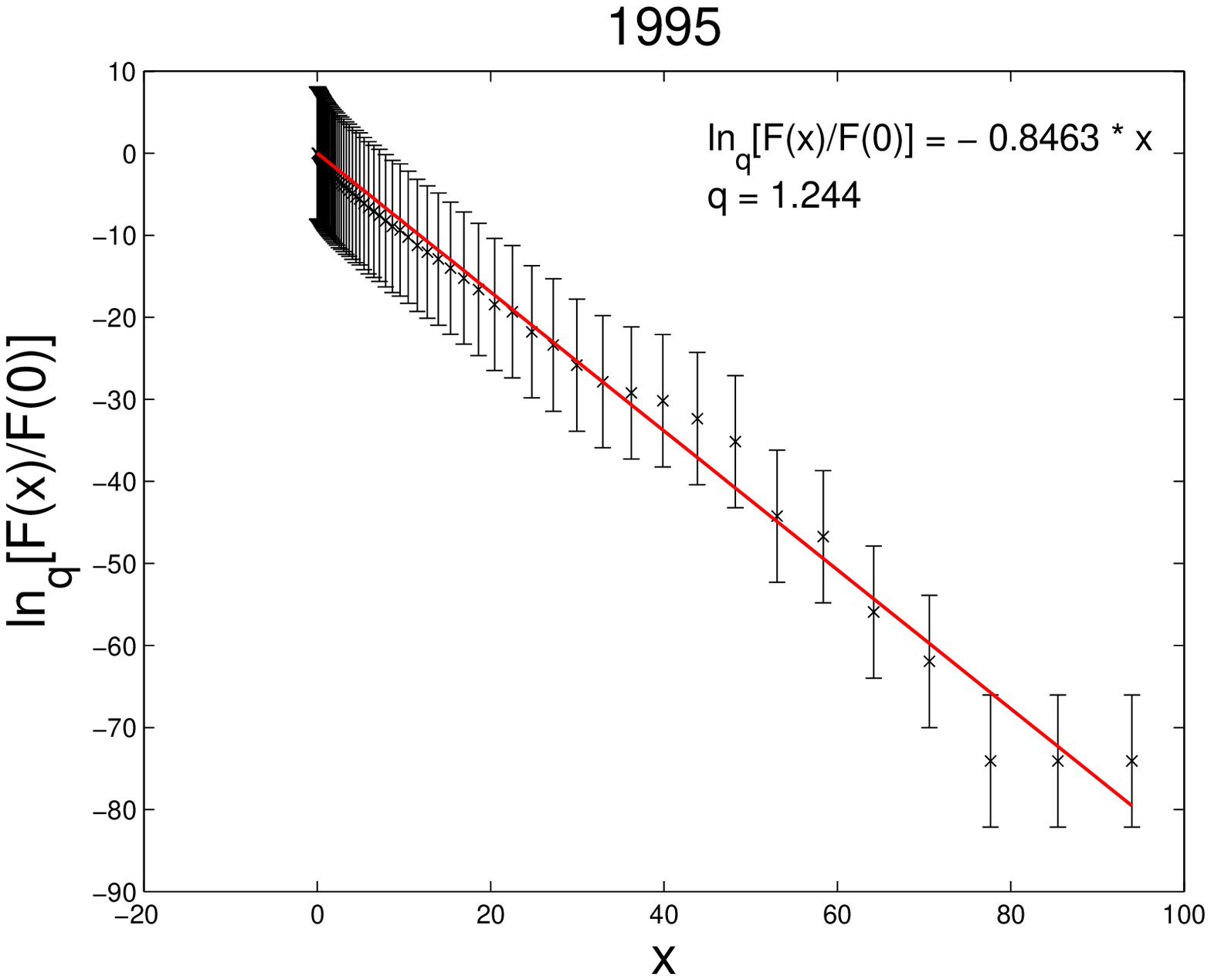} &
\includegraphics[scale=0.4]{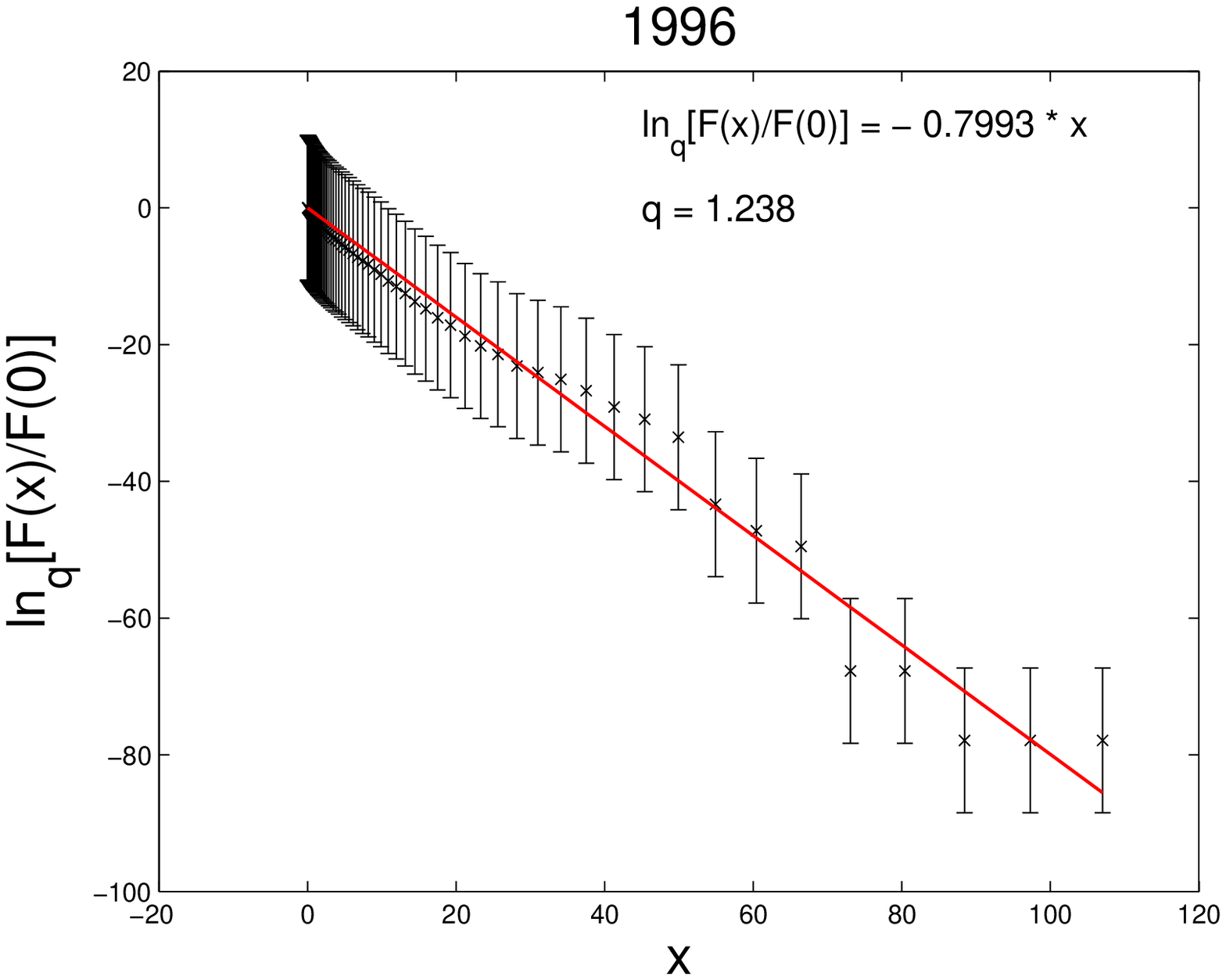} \\
\includegraphics[scale=0.4]{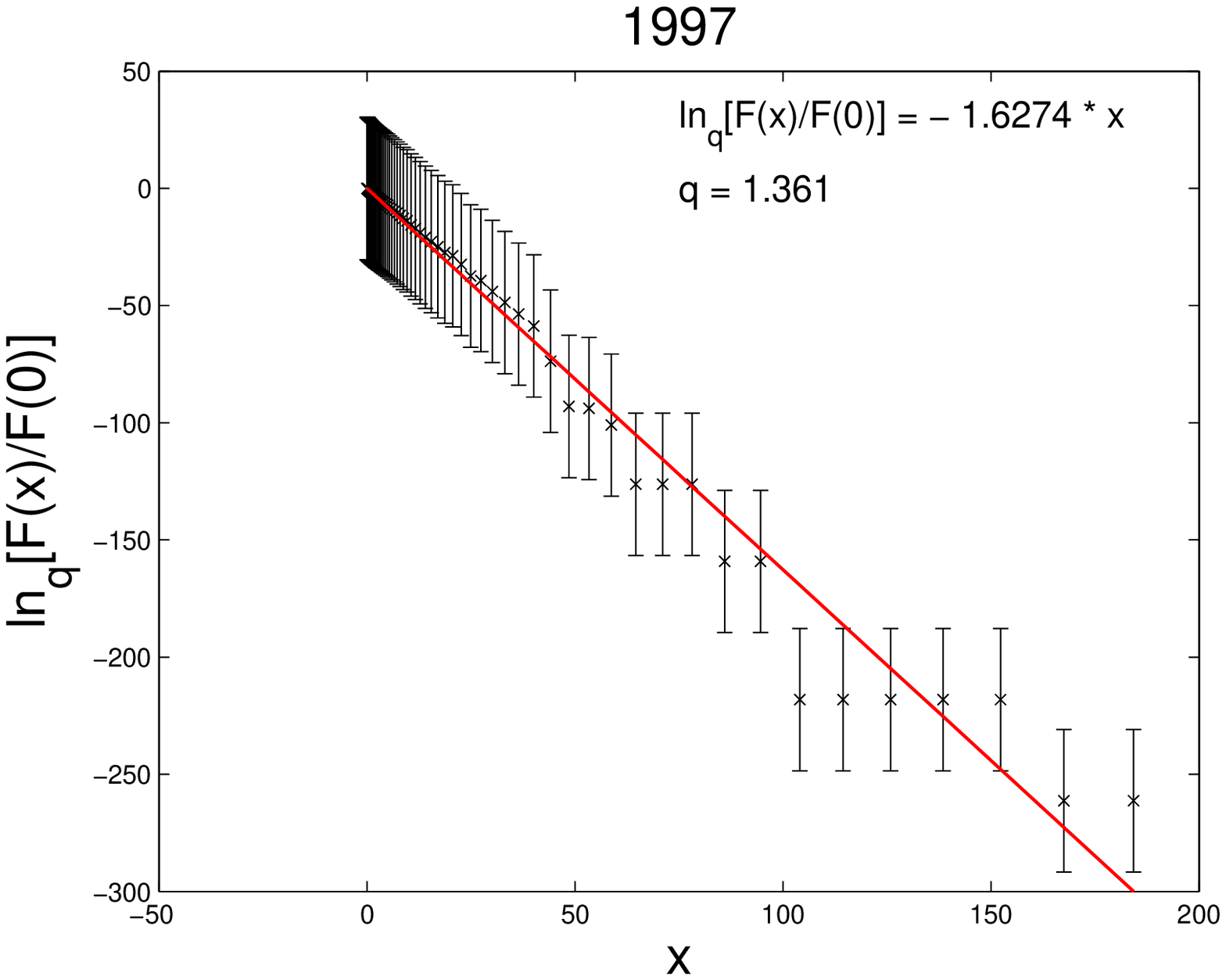} & 
\includegraphics[scale=0.4]{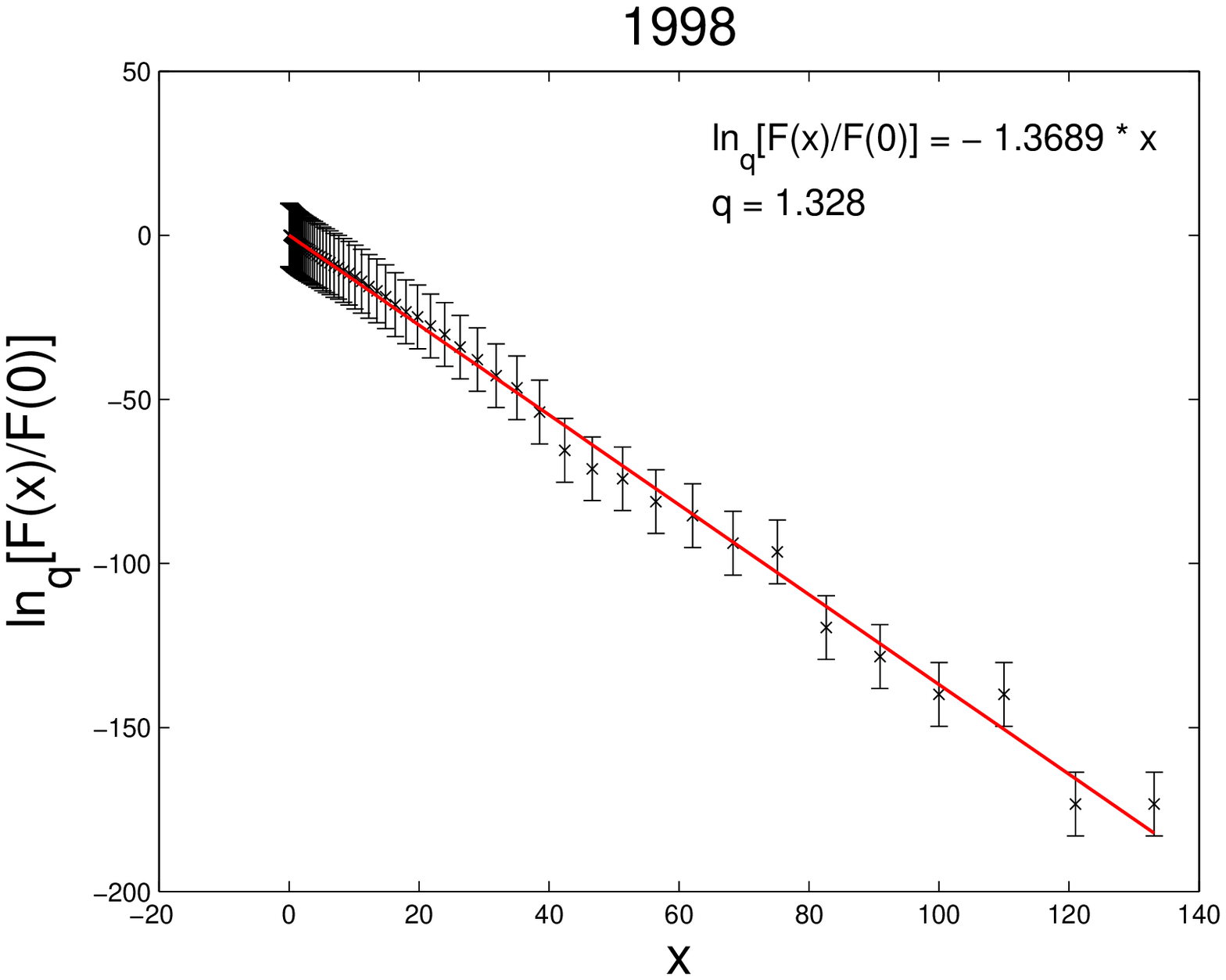} 
\end{array}$
\end{center}
\caption{Continuation of the previous graphs with data from 1992 to 1998.}
\lb{mosaic3}
\end{figure}

\begin{figure}[ht]
\begin{center}$
\begin{array}{ccc}
\includegraphics[scale=0.4]{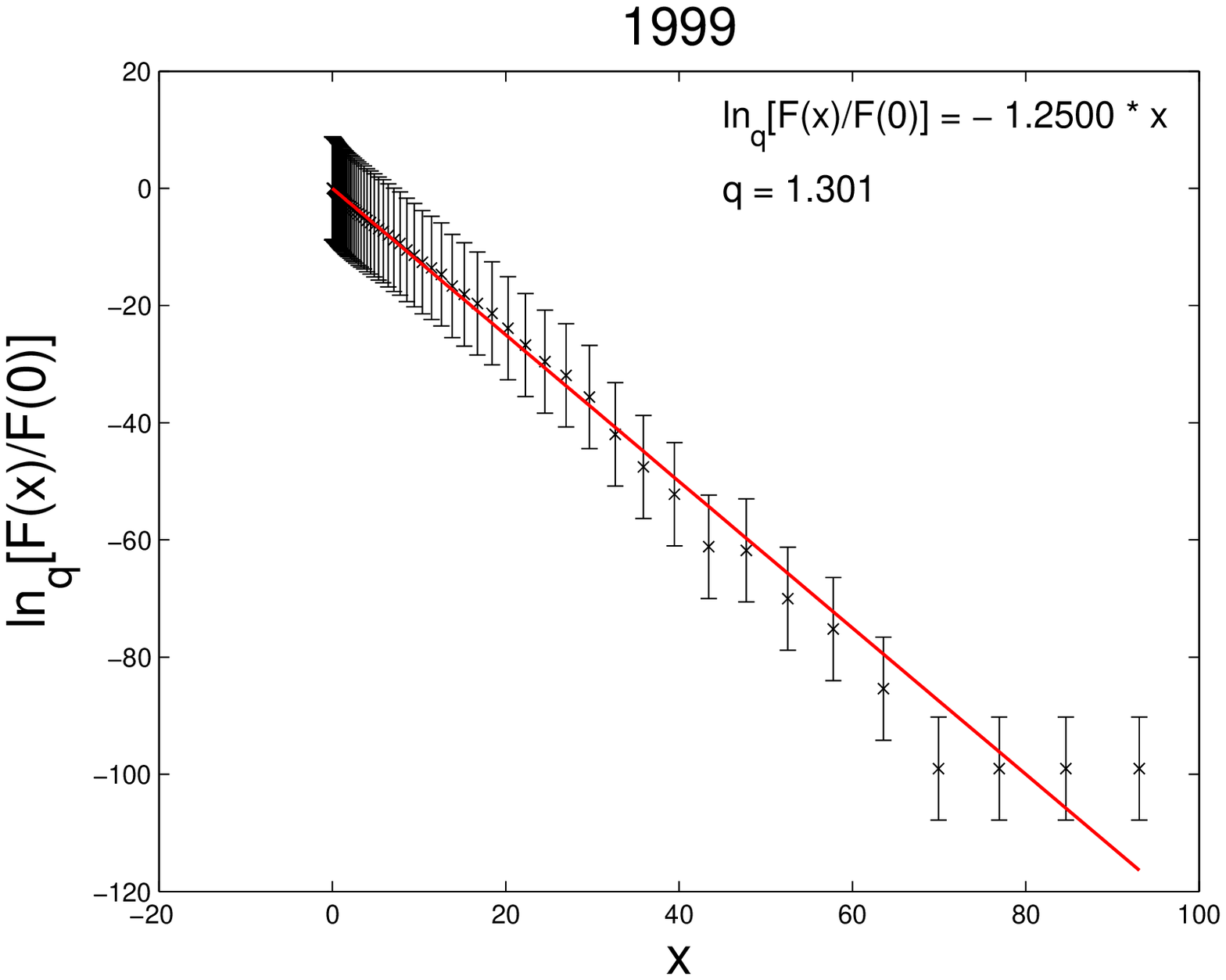} &
\includegraphics[scale=0.4]{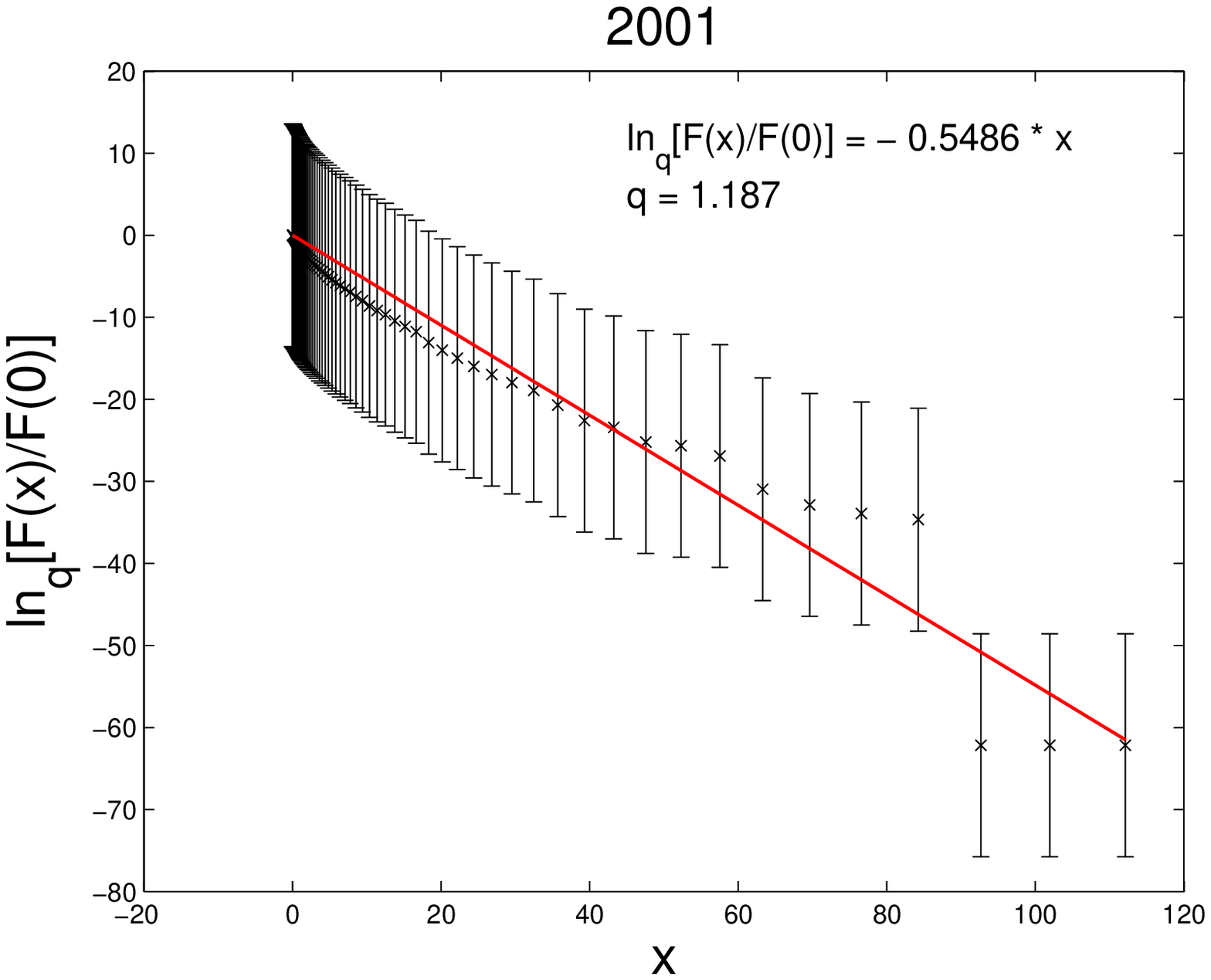} \\ 
\includegraphics[scale=0.4]{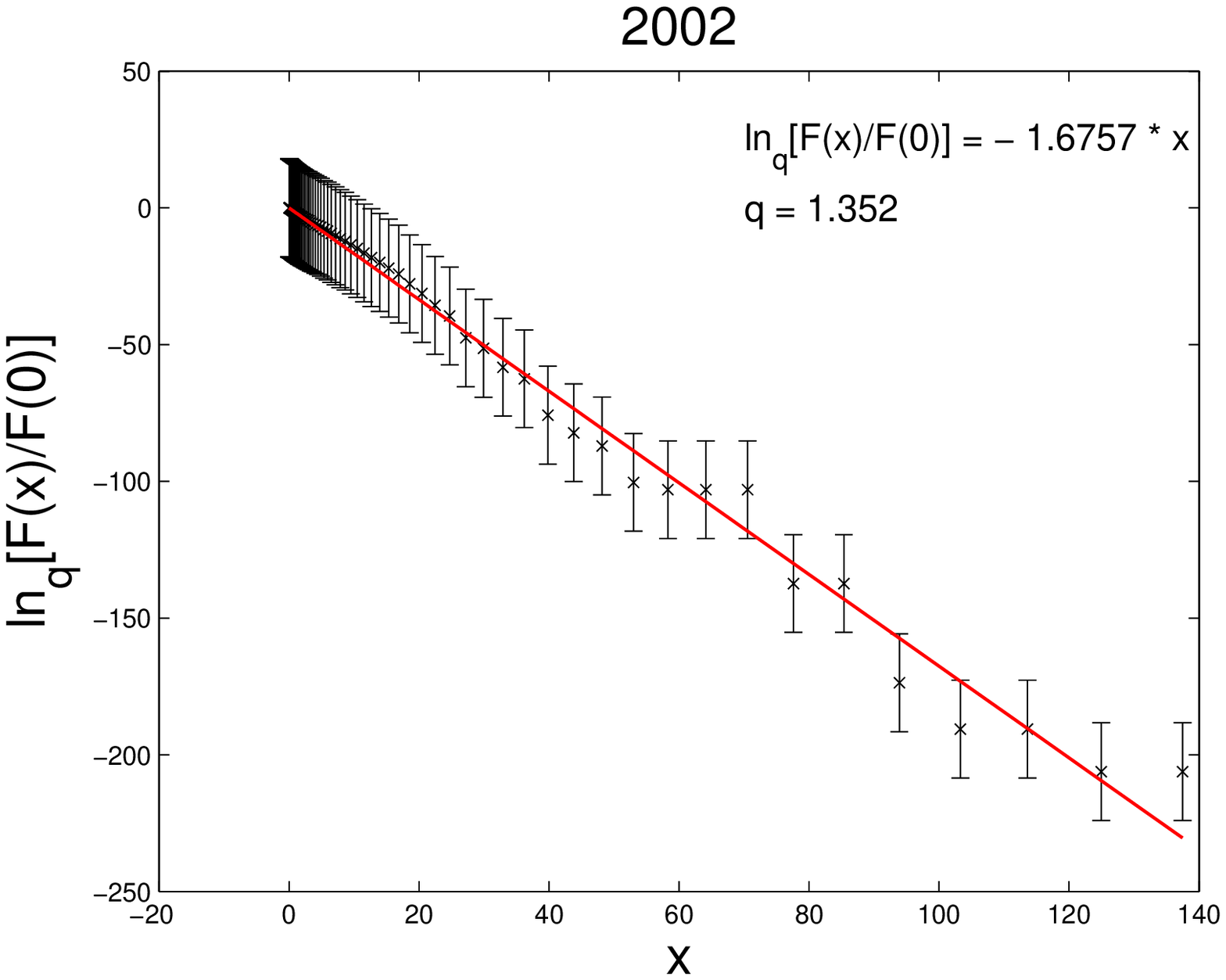} &
\includegraphics[scale=0.4]{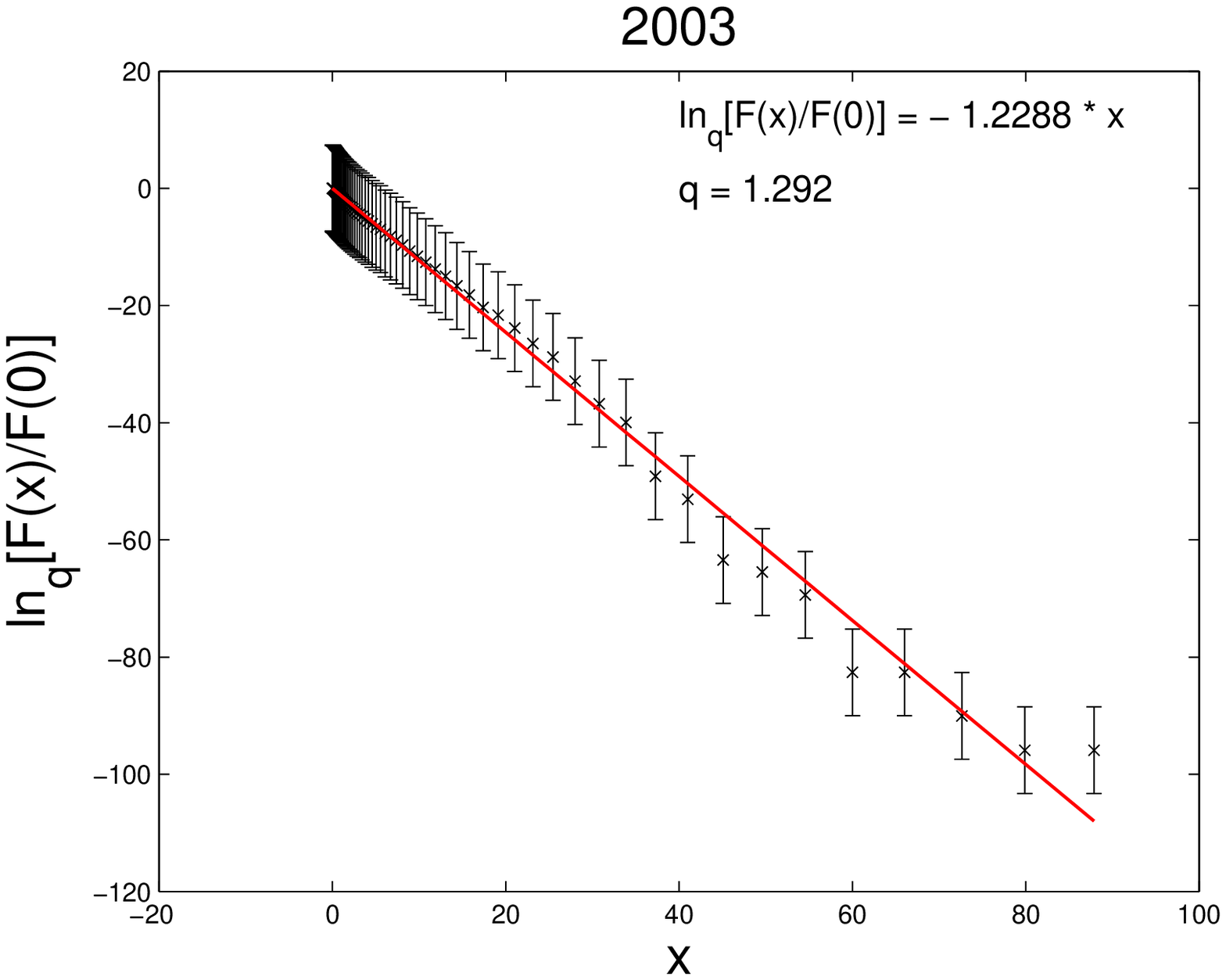} \\
\includegraphics[scale=0.4]{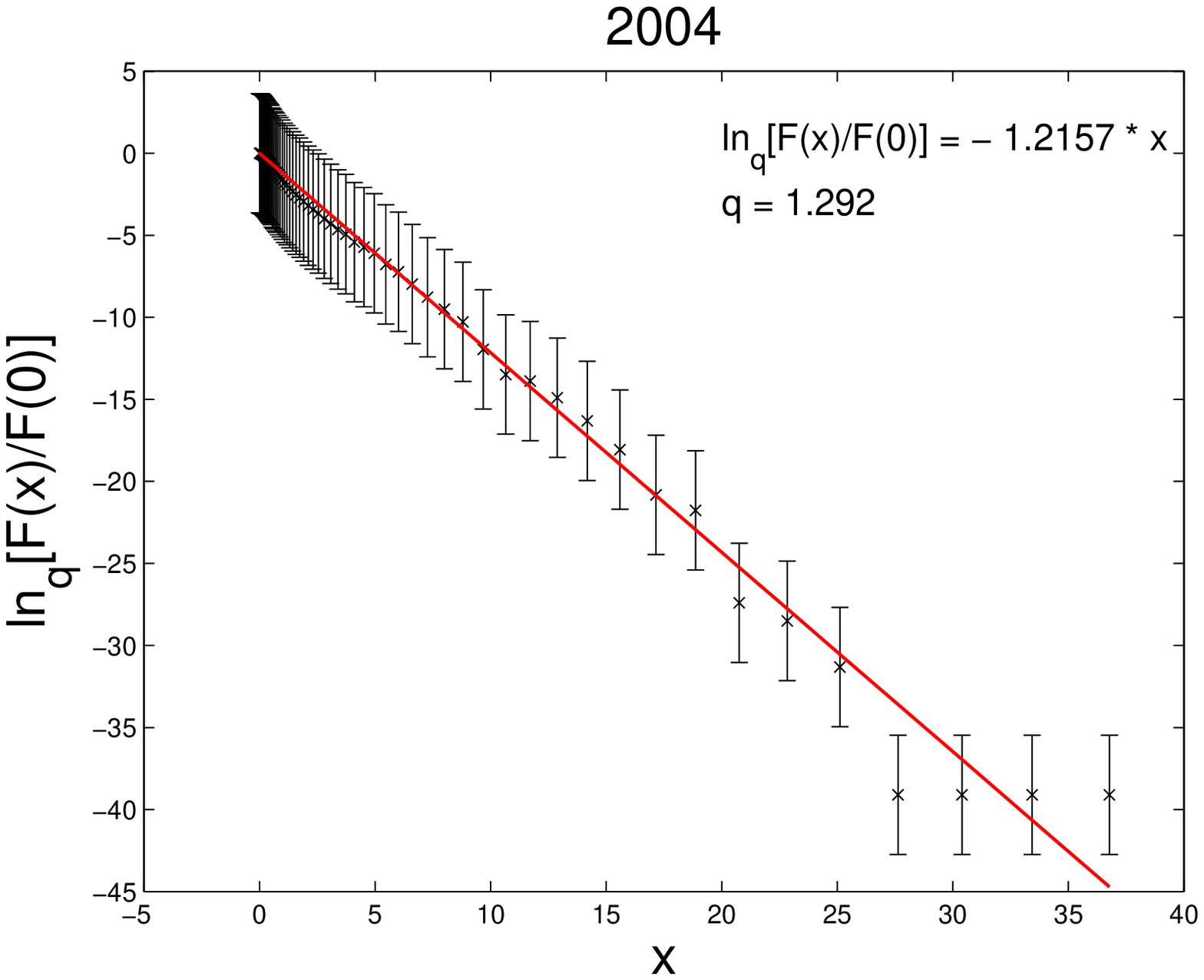} & 
\includegraphics[scale=0.4]{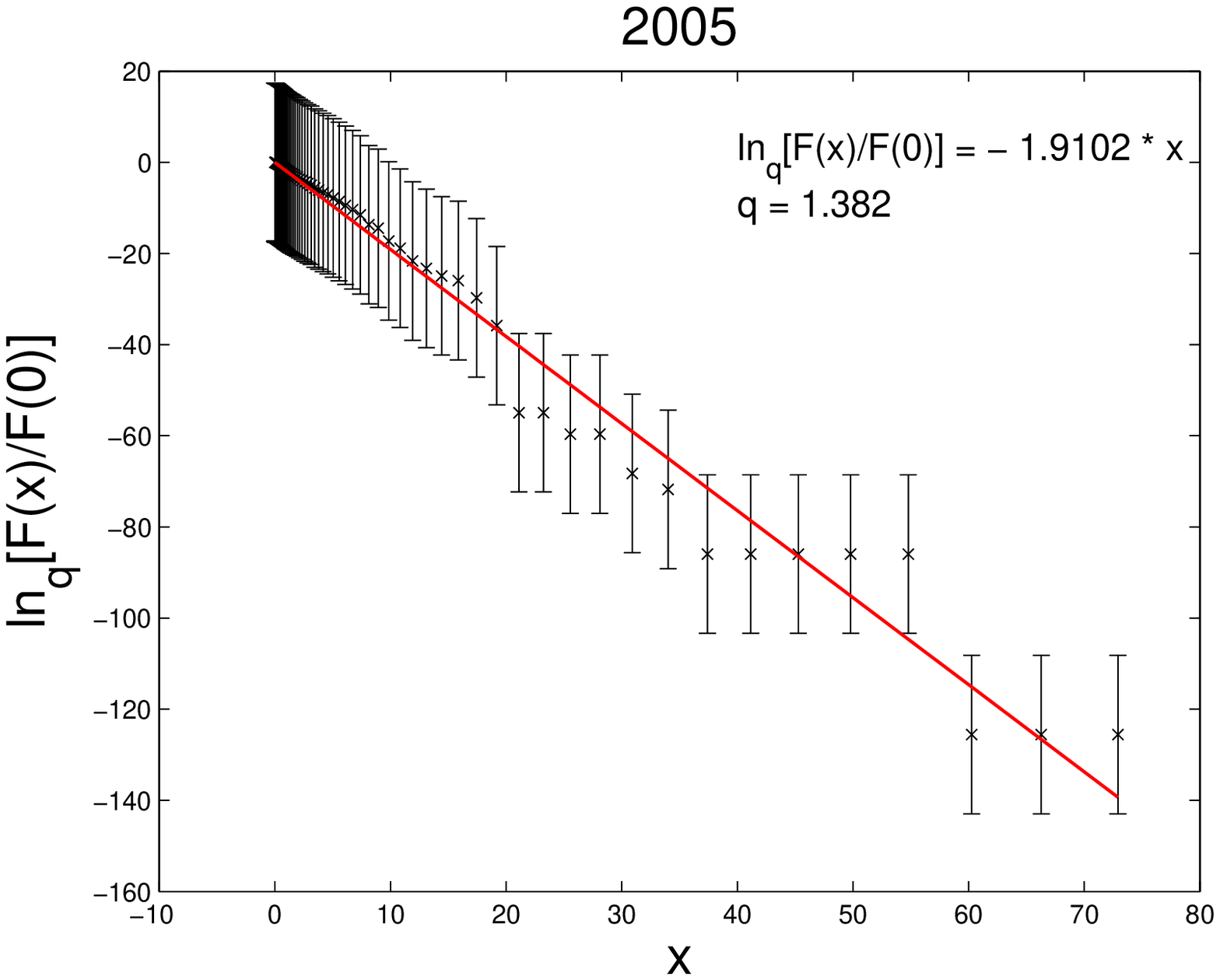} 
\end{array}$
\end{center}
\caption{Continuation of the previous graphs with data from 1999 to 2005.}
\lb{mosaic4}
\end{figure}

\begin{figure}[ht]
\begin{center}$
\begin{array}{cc}
\includegraphics[scale=0.4]{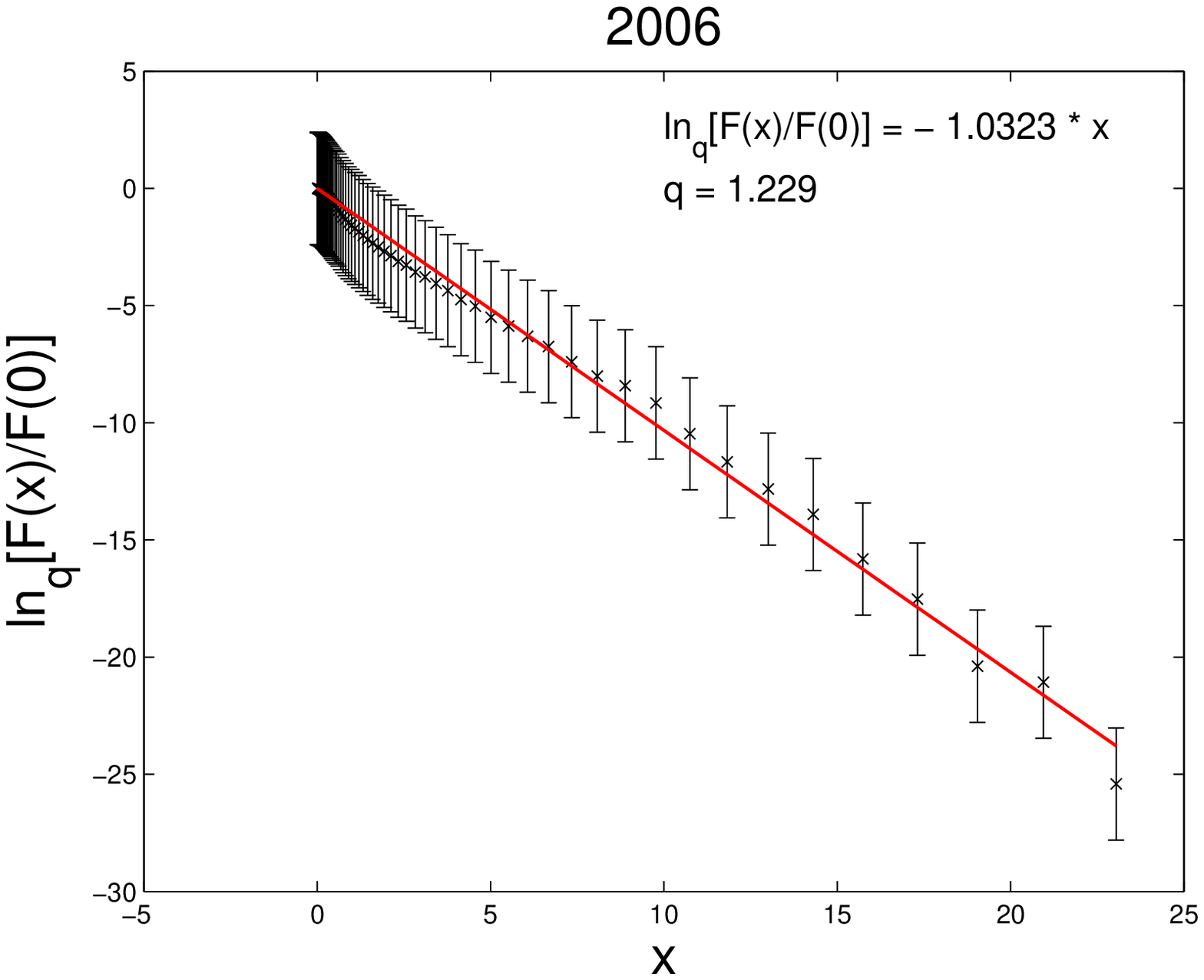} & 
\includegraphics[scale=0.4]{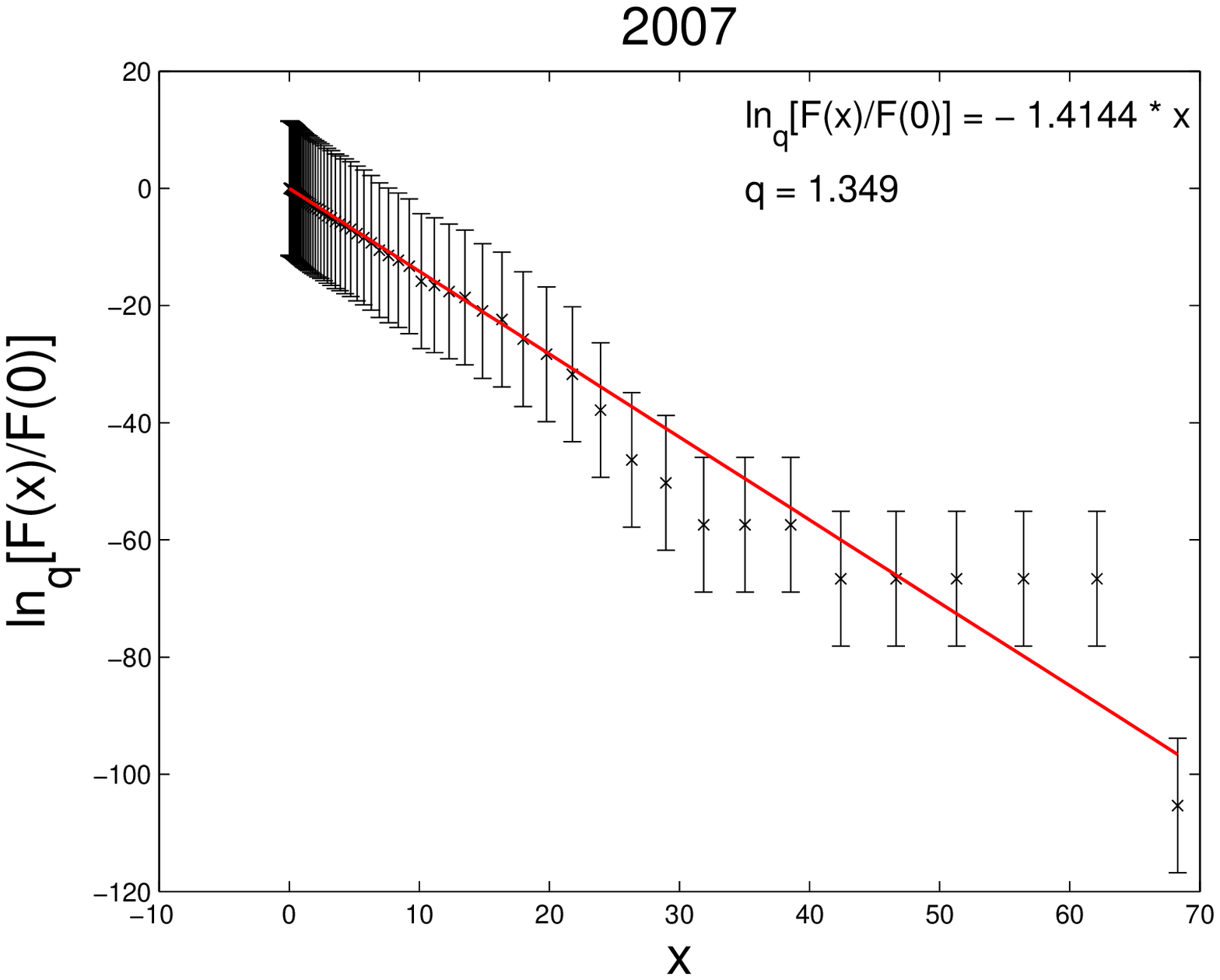} \\
\includegraphics[scale=0.4]{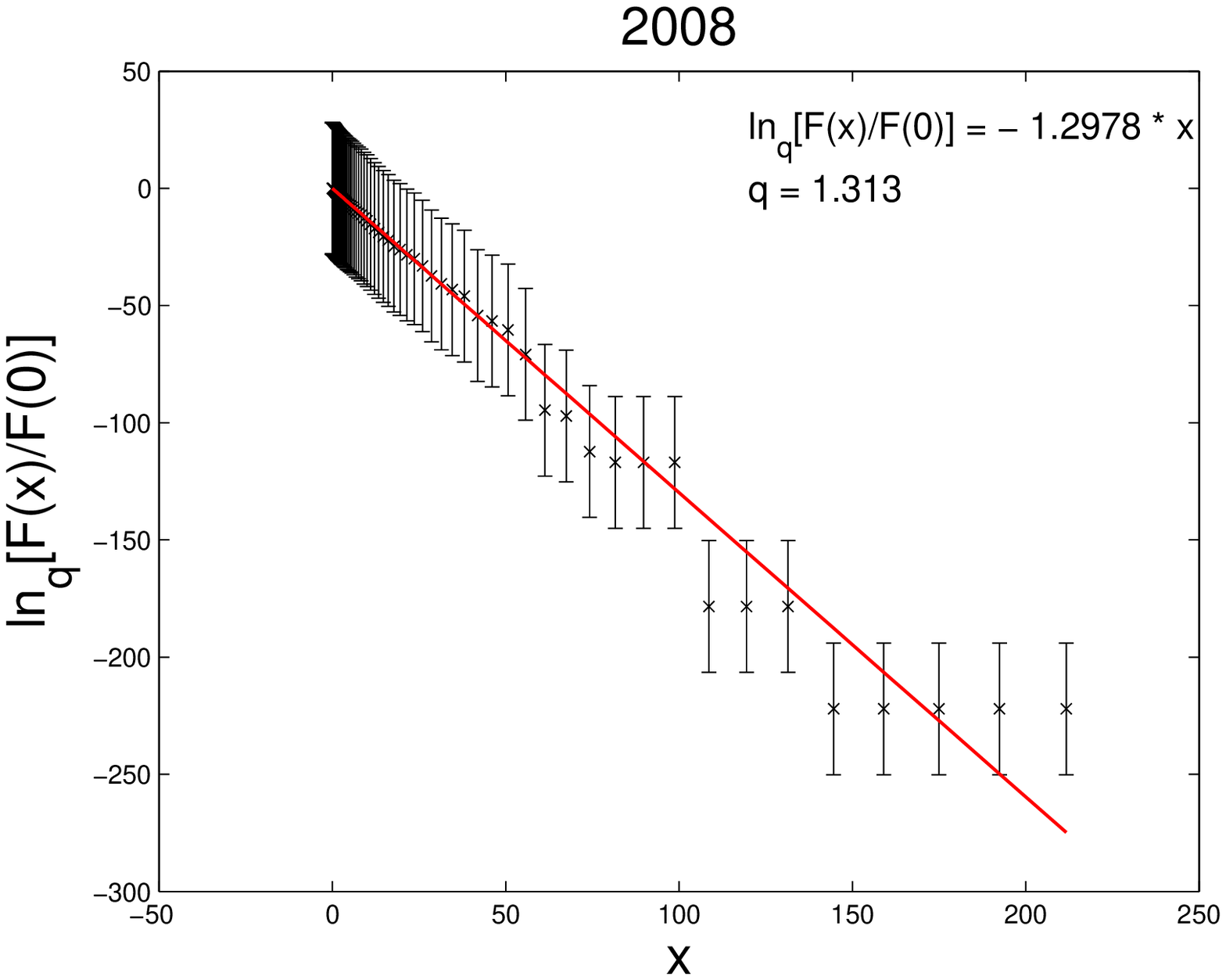} & 
\includegraphics[scale=0.4]{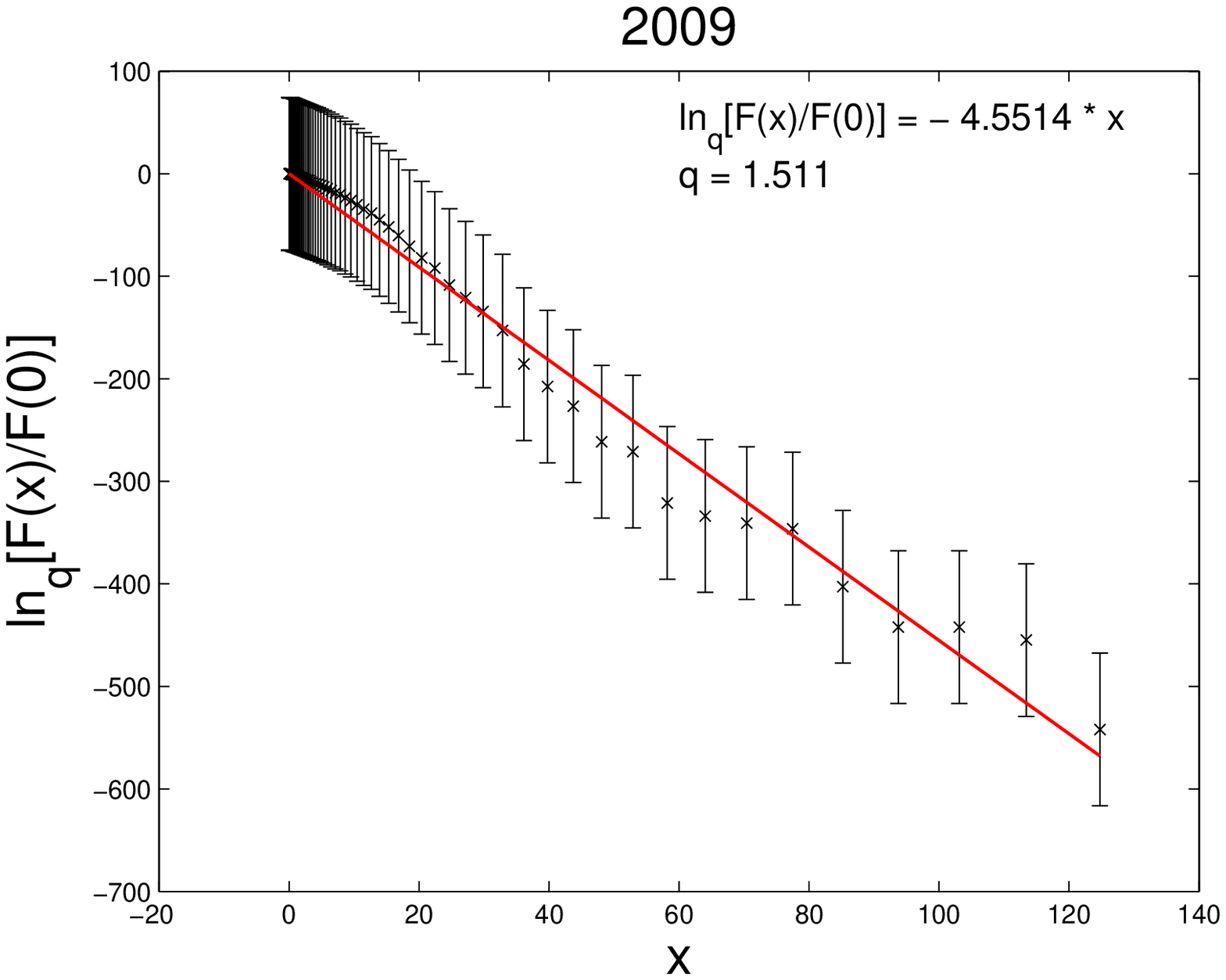} \\
\includegraphics[scale=0.4]{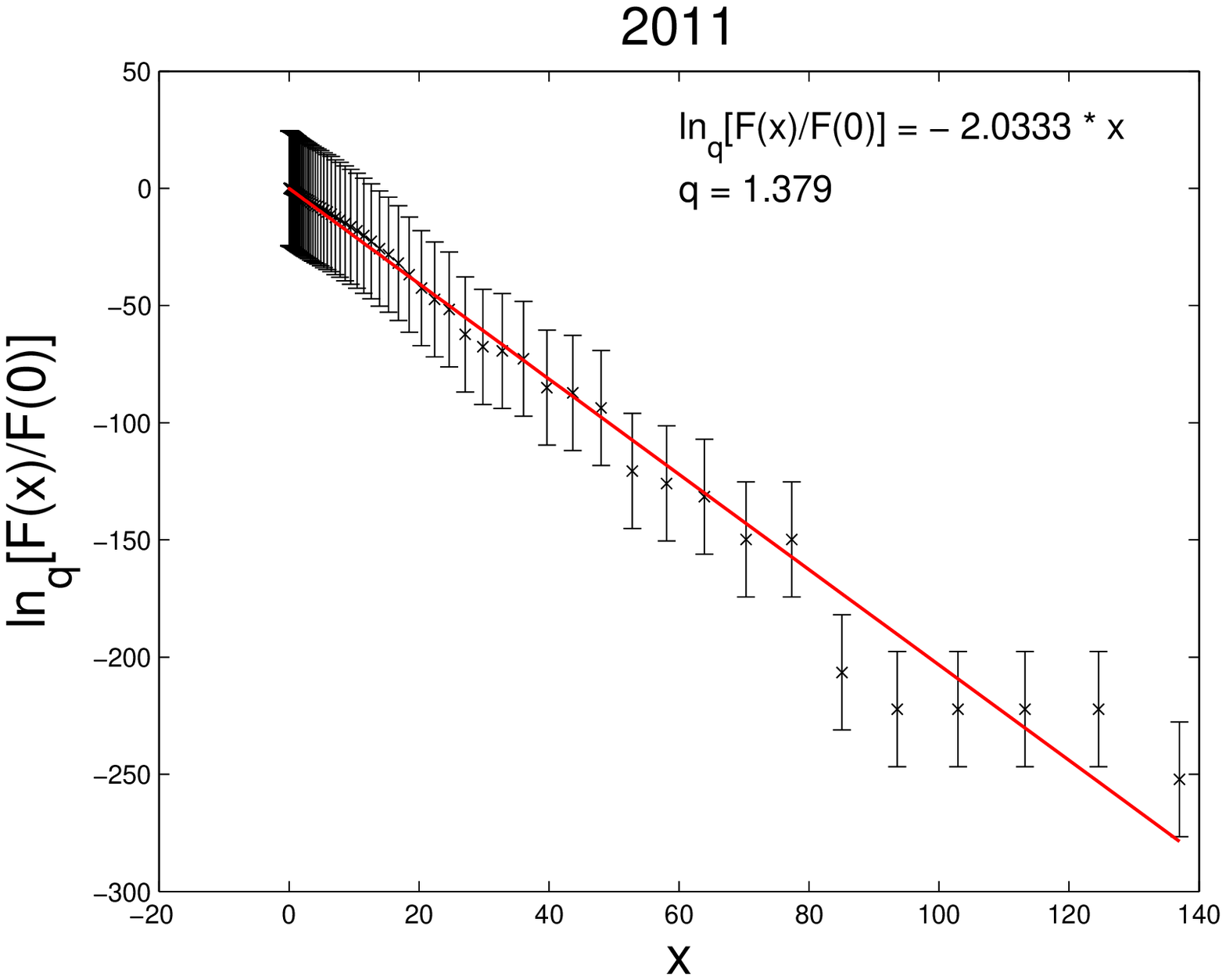} & 
\includegraphics[scale=0.4]{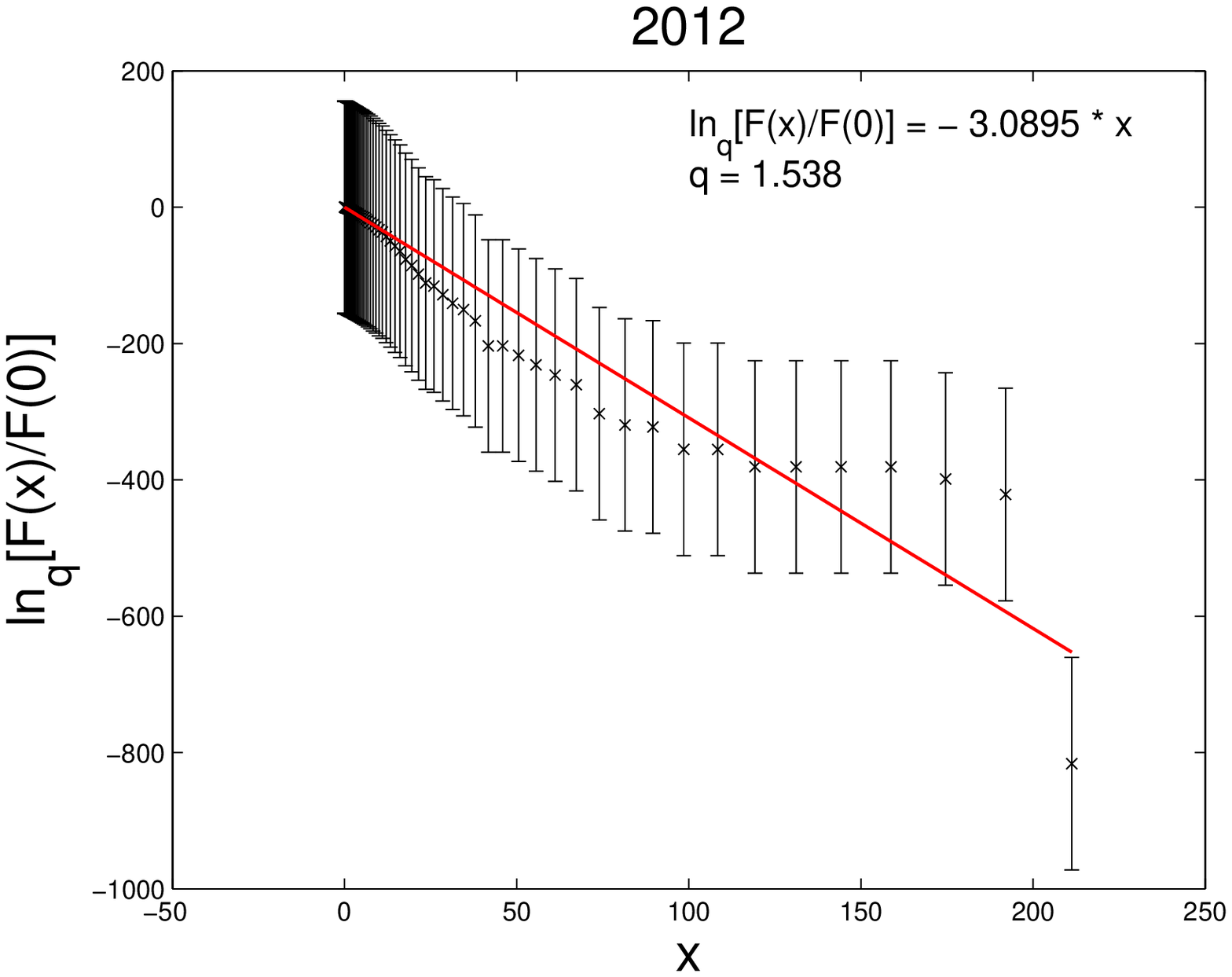} \\
\end{array}$
\end{center}
\caption{Continuation of the previous graphs with data from 2006 to 2012.}
\lb{mosaic5}
\end{figure}

\begin{figure}[ht]
\begin{center}$
\begin{array}{cc}
\includegraphics[scale=0.4]{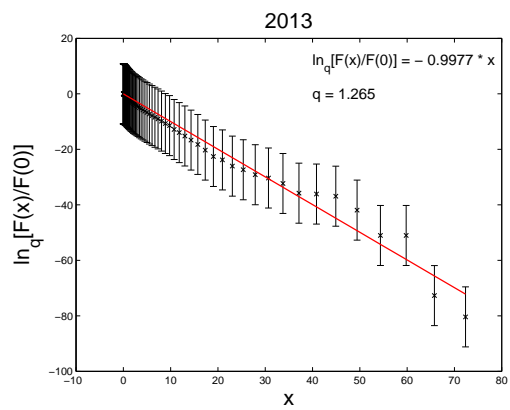} & 
\includegraphics[scale=0.4]{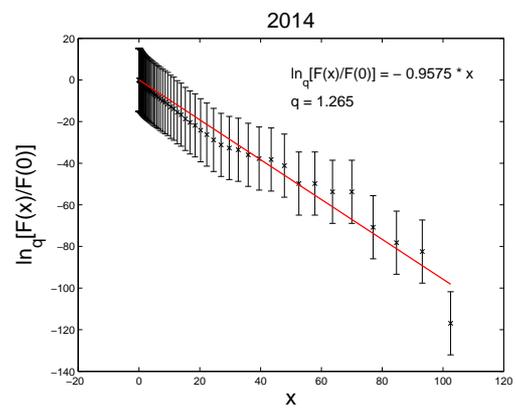} \\
\end{array}$
\end{center}
\caption{Continuation of the previous graphs with data for 2013 and 2014.}
\lb{mosaic6}
\end{figure}

\begin{figure}[ht]
\begin{center}$
\begin{array}{cc}
\includegraphics[scale=0.5,angle=-90]{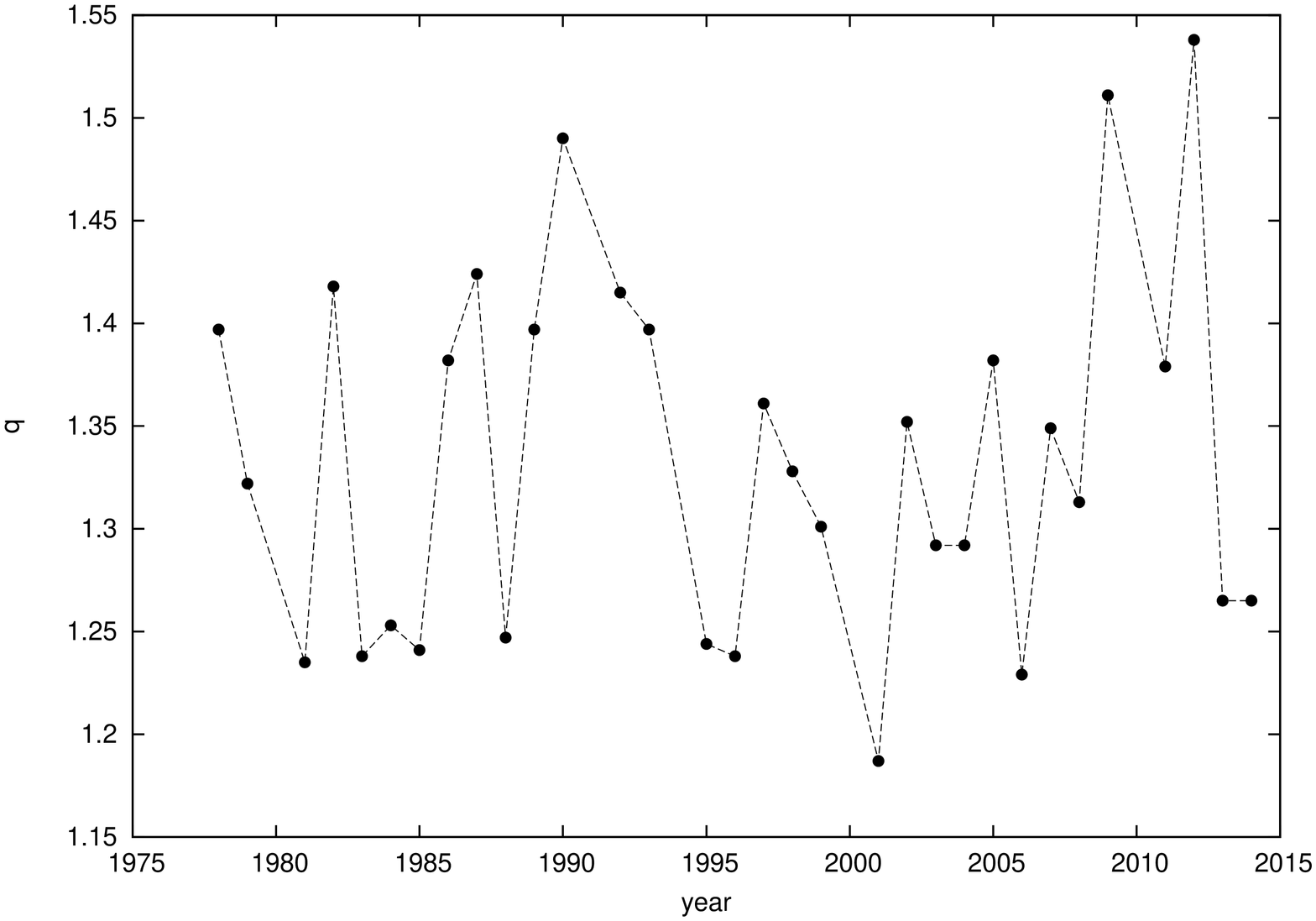} \\ 
\includegraphics[scale=0.5,angle=-90]{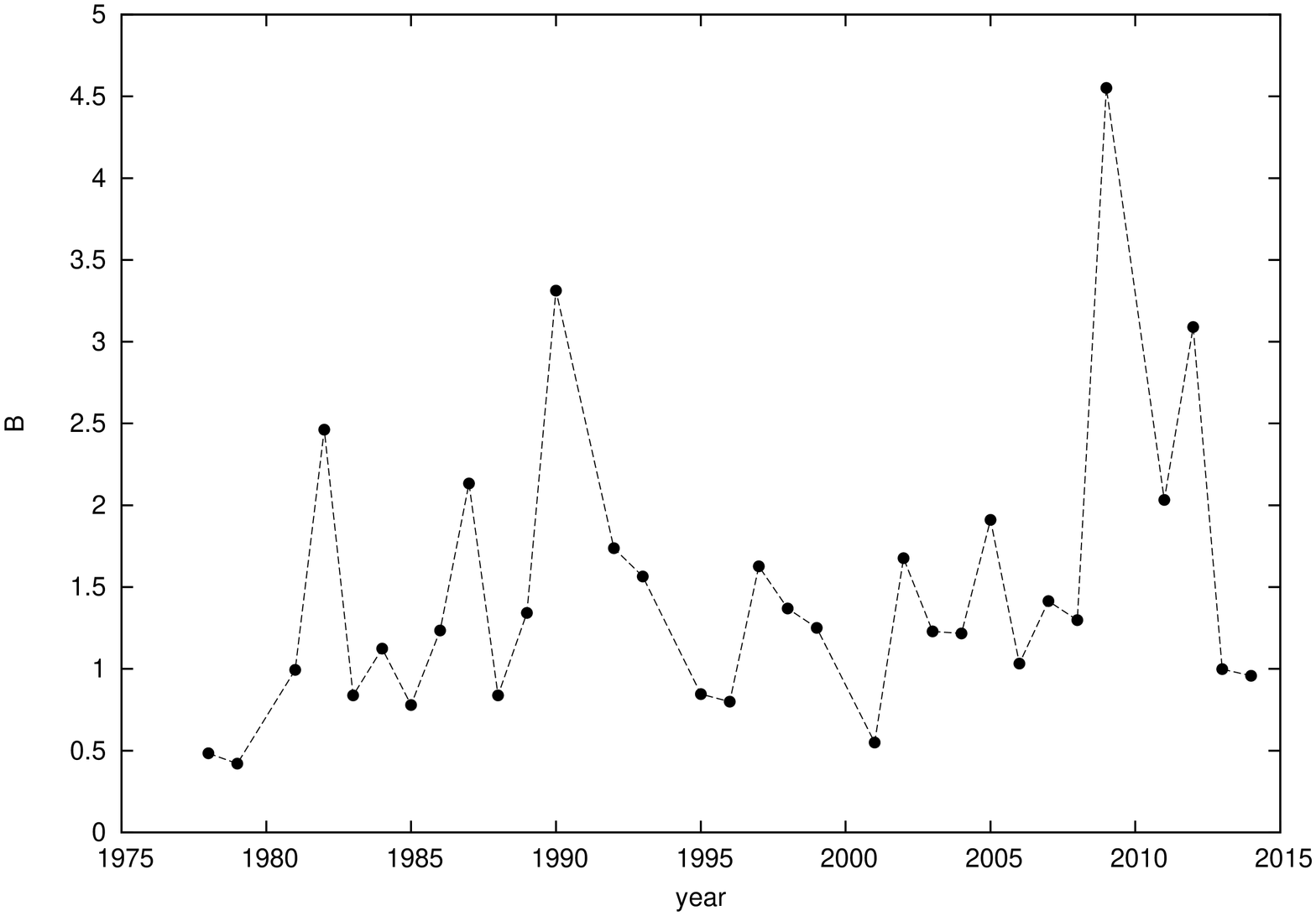} 
\end{array}$
\end{center}
\caption{Plots of the parameters $q$ (top) and $B$ (bottom) in terms of
the time span of the samples in years as given in Table \ref{tab1}. It
is clear that both parameters oscillate periodically with maxima from
2 to 5 years interval. The oscillation period is about 3.5 years on
average and the maxima and minima of both $B$ and $q$ mostly coincide,
a fact which suggests a pattern between them (see Fig.\ \ref{B-q} below).}
\lb{q-B-year}
\end{figure}

\begin{figure}[ht]
\begin{center}$
\includegraphics[scale=0.5,angle=-90]{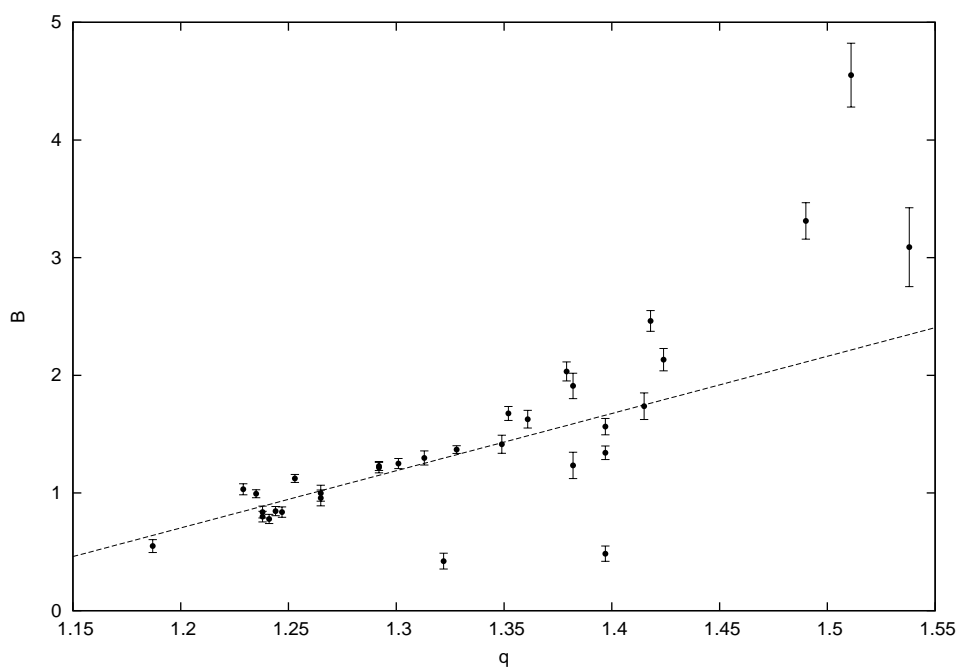}
$  
\end{center}
\caption{Graph of $B$ vs.\ $q$ with error bars in $B$. The plot
suggests the existence of a linear growth pattern between the two
parameters. Although labels indicating the correspondent year of
each point were omitted to avoid image clutter, this growth pattern
seems unrelated to time evolution. The dashed line is a weighted
(in $B$) linear fit to the points in the form $B=aq+b$, having the
following fitted parameters: $a=4.86\pm0.88$ and $b=-5.13\pm1.13$.}
\lb{B-q}
\end{figure}

\begin{figure}[ht]
\begin{center}$
\begin{array}{cc}
\includegraphics[scale=0.4,angle=-90]{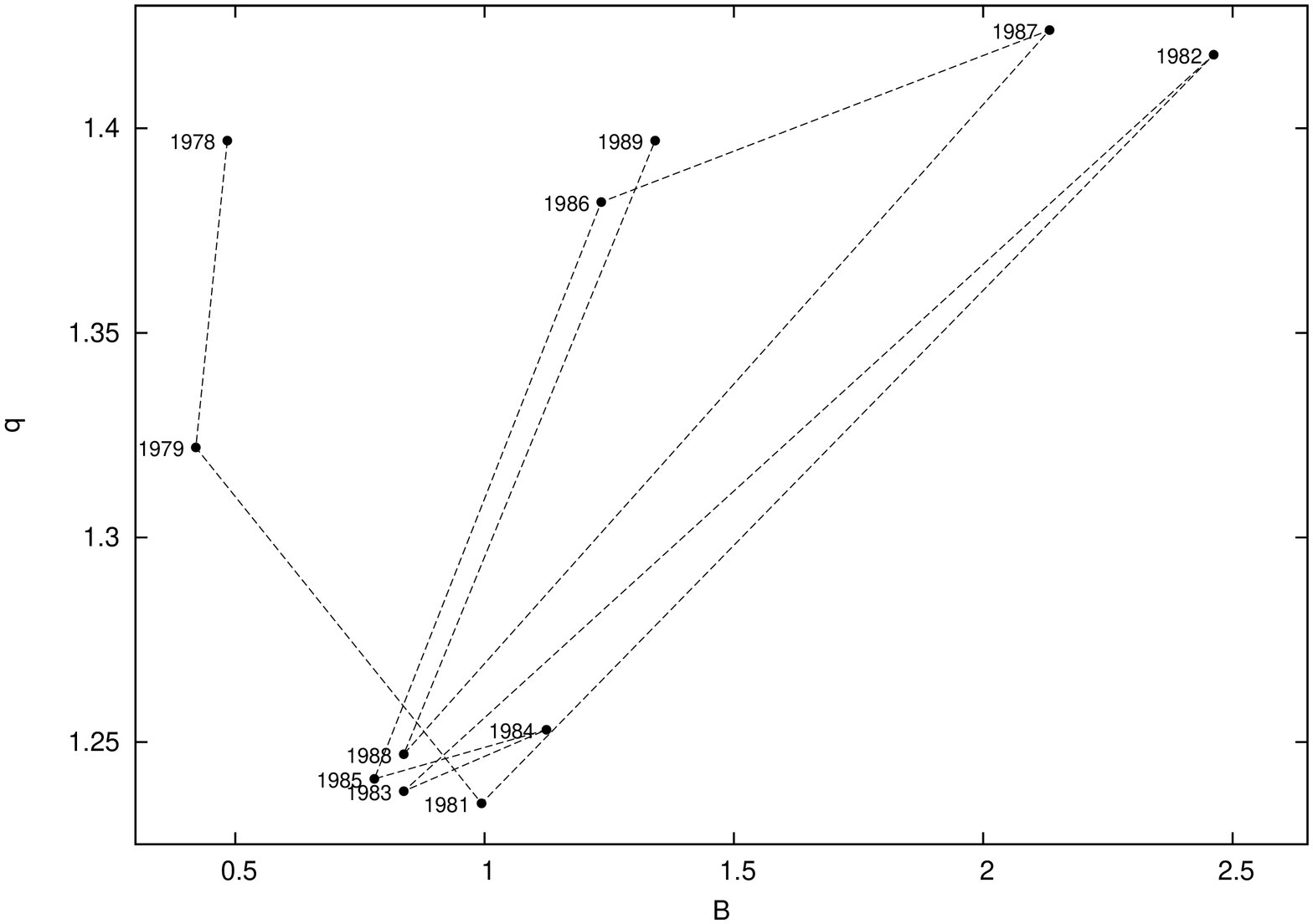} \\ 
\includegraphics[scale=0.4,angle=-90]{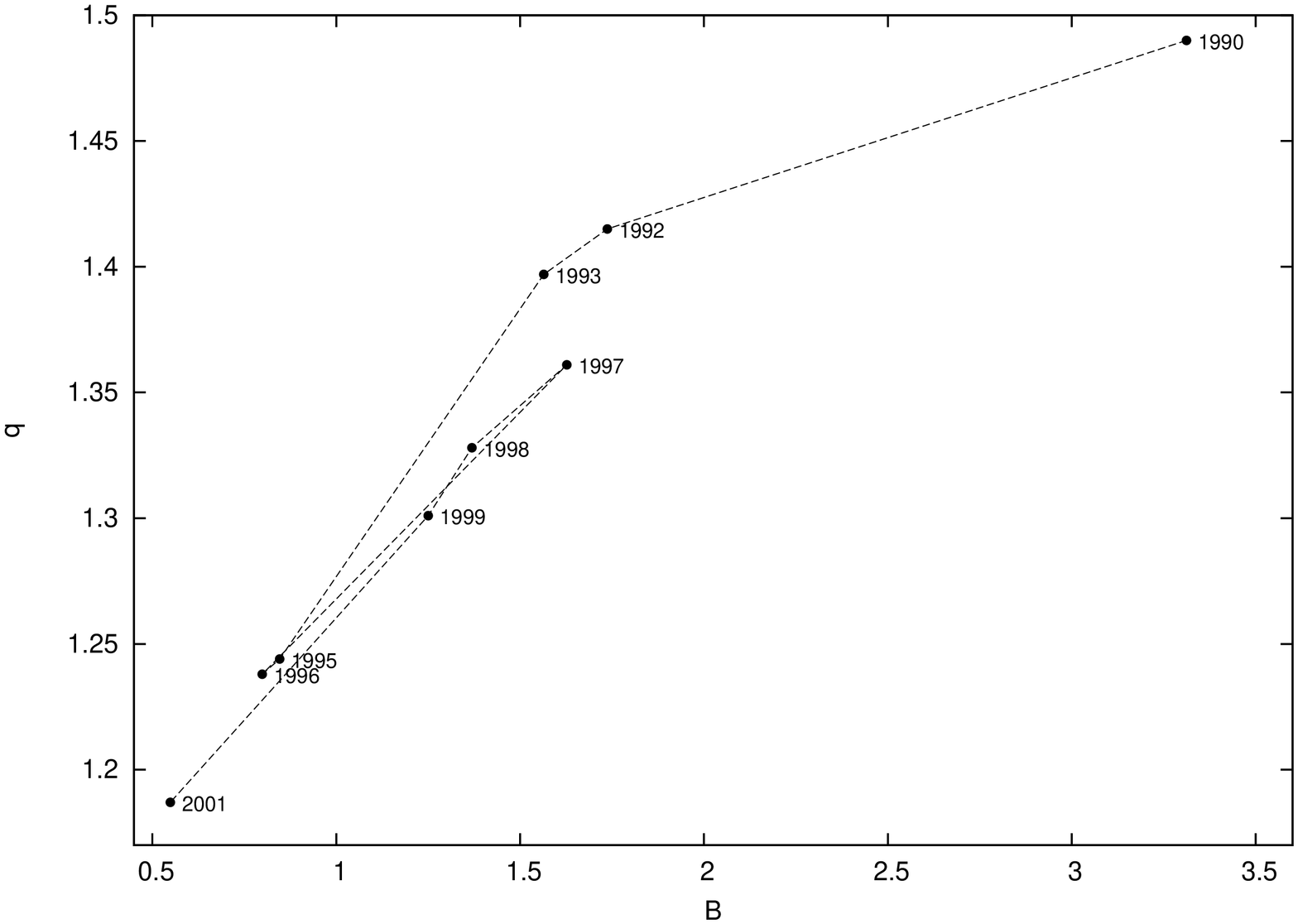} \\
\includegraphics[scale=0.4,angle=-90]{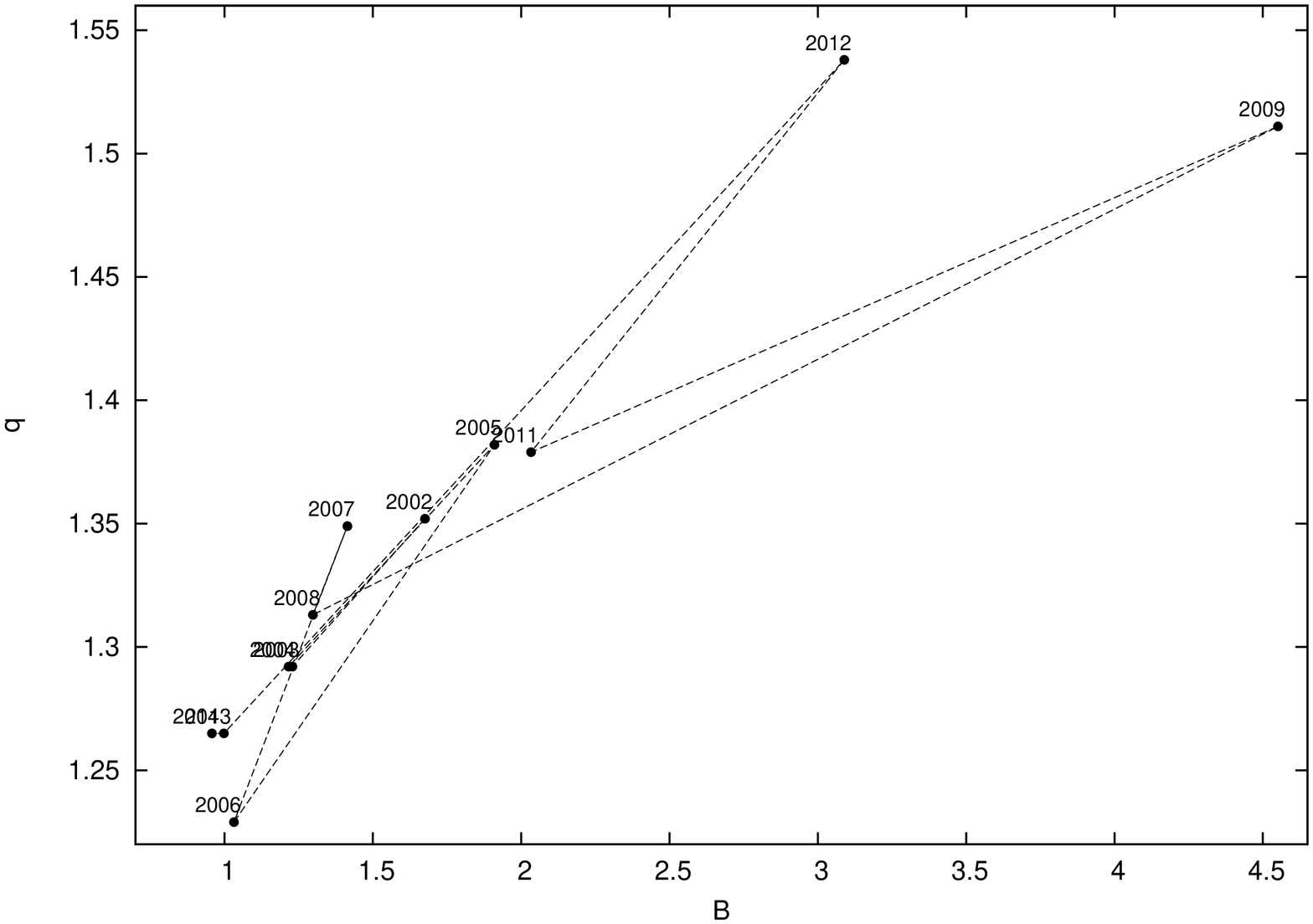} 
\end{array}$
\end{center}
\caption{These graphs are a different way of representing the points
shown in Fig.\ \ref{B-q} above. Here the axes are inverted, showing a
$q$ vs.\ $B$ plane, and, to avoid image clutter, the time interval was
broken in three segments, from 1978 to 1989 (top), 1990 to 2001 (middle)
and 2002 to 2014 (bottom). The dashed lines connect the points
chronologically and a pattern appear in the form of a general clockwise
cycle without a single center in all plots, but having a few
anti-clockwise turns. On the bottom graph, the points representing the
years 2003 and 2004 are close enough to be almost superimposed. The same
happens to 2013 and 2014.}
\lb{q-B-ciclos}
\end{figure}

\begin{figure}[ht]
\begin{center}$
\includegraphics[scale=0.55]{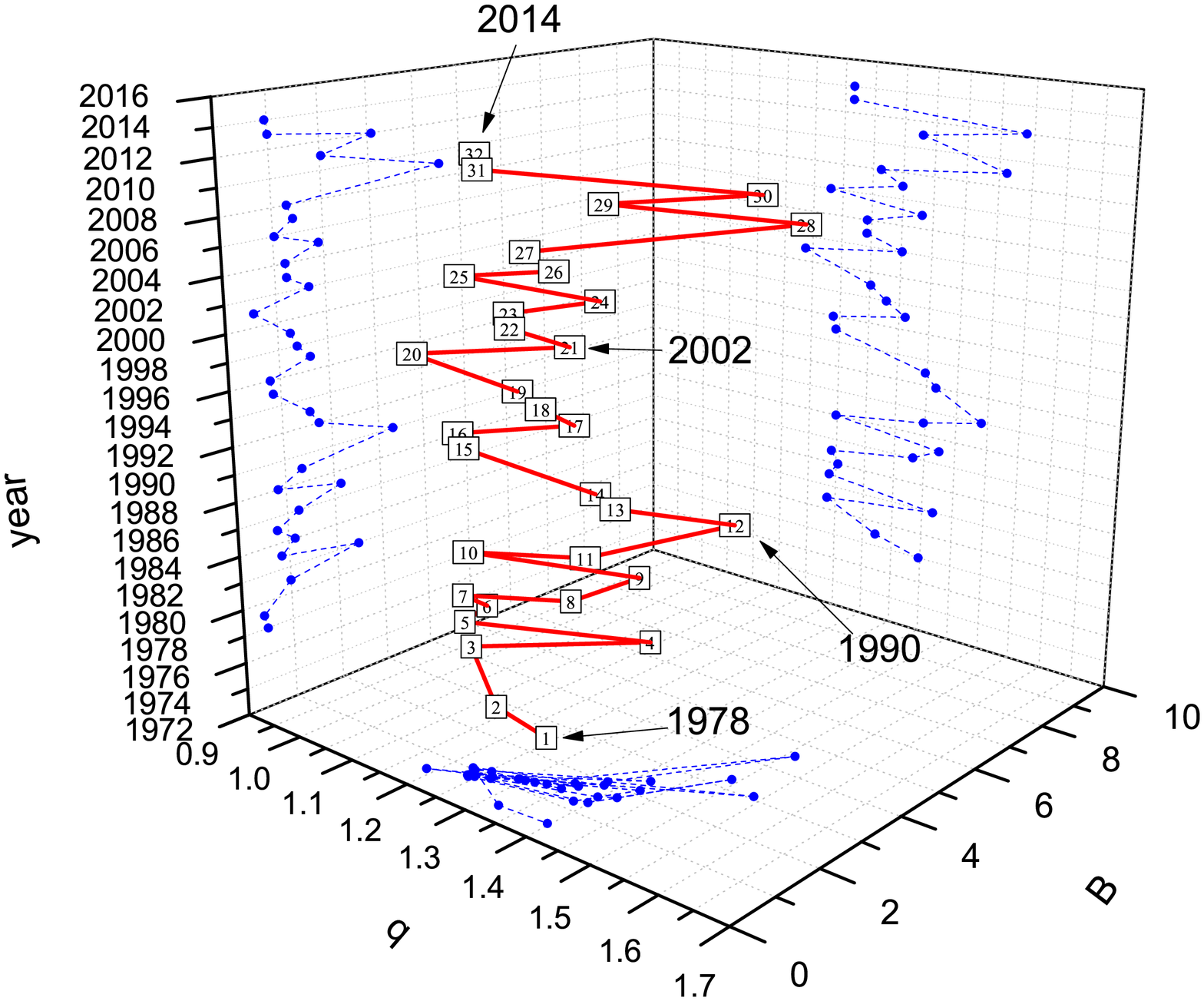}
$  
\end{center}
\caption{Three dimensional plot of $q$ vs.\ $B$ vs.\ year that summarizes
the results of Figs.\ \ref{q-B-year}--\ref{q-B-ciclos} in
addition to showing a helix type line evolving mostly clockwise at
increasing time in the $z$ axis, according to the left hand rule. All
points are numbered in ascending chronological order and connected by a
red line in order to make the helical like shape evolution visible. The
points representing the years 1978, 1990, 2002 and 2014 are explicitly
indicated for additional clarity. The graphs of Fig.\ \ref{q-B-year}
appear in blue and in inverted order on the $xz$ and $yz$ planes. All plots
of Fig.\ \ref{q-B-ciclos} appear together projected in the $xy$ plane,
where one can also distinguish the growth pattern discussed in Fig.\
\ref{B-q}.}
\lb{B-q-year-3D.eps}
\end{figure}

\vspace{4mm}
\small
Our thanks go to C.\ Tsallis for the initial suggestions which led
to this paper. We are also grateful to two referees for pointing
out relevant literature on the log-periodic oscillations and useful
comments and suggestions. One of us (M.B.R.) acknowledges partial financial
support from FAPERJ.
\vspace{4mm}


\begin{thebibliography}{99}
\bibitem{pareto}V.\ Pareto, \textit{``Cours d'\'{E}conomie Politique''},
   Lausanne, 1897
\bibitem{k80}N.C.\ Kakwani, \textit{``Income Inequality and Poverty''},
   Oxford University Press, 1980
\bibitem{nm09}N.J.\ Moura Jr., M.B.\ Ribeiro, \textit{``Evidence for
   the Gompertz Curve in the Income Distribution of Brazil 1978-2005''},
   Eur.\ Phys.\ J.\ B, 67 (2009) 101-120, arXiv:0812.2664v1
\bibitem{fnm10}F.\ Chami Figueira, N.J.\ Moura Jr., M.B.\ Ribeiro,
   \textit{``The Gompertz-Pareto Income Distribution''}, Physica A,
   390 (2011) 689-698, arXiv:1010.1994v1 
\bibitem{dy01}A.\ Dr\u{a}gulescu, V.M.\ Yakovenko, \textit{``Evidence
   for the Exponential Distribution of Income in the USA''}, Eur.\ Phys.\
   J.\ B, 20 (2001) 585, arXiv:cond-mat/0008305v2 
\bibitem{ferrero2} J.C.\ Ferrero, \textit{``The Statistical Distribution
    of Money and the Rate of Money Transference''}, Physica A, 341 (2004)
    575-585
\bibitem{s05}A.\ Christian Silva, \textit{``Applications of Physics to
   Finance and Economics: Returns, Trading Activity and Income''}, PhD
   thesis, University of Maryland, 2005, arXiv:physics/0507022v1
\bibitem{yr09} V.M.\ Yakovenko, J.B.\ Rosser, \textit{``Colloquium:
    Statistical Mechanics of Money, Wealth, and Income''}, Rev.\ Mod.\
    Phys., 81 (2009) 1703-1725, arXiv:0905.1518v2
\bibitem{crbh08} R.\ Coelho, P.\ Richmond, J.\ Barry, S.\ Hutzler,
   \textit{``Double Power Laws in Income and Wealth Distributions''},
   Physica A, 387 (2008) 3847-3851, arXiv:0710.0917v1
\bibitem{by10} A.\ Banerjee, V.M.\ Yakovenko, \textit{``Universal
   Patterns of Inequality''}, New J.\ Phys., 12 (2010) 075032,
   arXiv:0912.4898v4
\bibitem{nm13} N.J.\ Moura Jr., M.B.\ Ribeiro, \textit{``Testing the
   Goodwin Growth-Cycle Macroeconomic Dynamics in Brazil''}, Physica A,
   392 (2013) 2088-2103, arXiv:1301.1090 
\bibitem{cyc05} Econophys-Kolkata I Workshop, \textit{``Econophysics of
	Wealth Distributions''}, A.\ Chatterjee, S.\ Yarlagadda, B.K.\
	Chakrabarti (Eds.), Springer, 2005
\bibitem{c5} B.K.\ Chakrabarti, A.\ Chakraborti, S.R.\ Chakravarty, 
   A.\ Chatterjee, \textit{``Econophysics of Income and Wealth
   Distributions''}, Cambridge University Press, 2013
\bibitem{borges} E.P.\ Borges, \textit{``Empirical nonextensive laws for
    the county distribution of total personal income and gross domestic
    product''}, Physica A, 334 (2004) 255-266
\bibitem{ferrero1} J.C.\ Ferrero, \textit{``The Monomodal, Polymodal,
    Equilibrium and Nonequilibrium Distribution of Money''}, In
    \cite{cyc05}, pp.\ 159-167, (2005)
\bibitem{ferrero3} J.C.\ Ferrero, \textit{``An Statistical Analysis of
    Stratification and Inequality in the Income Distribution''}, Eur.\
    Phys.\ J.\ B 80 (2011) 255-261
\bibitem{tsallis1} C.\ Tsallis, \textit{``What are the Numbers that
	Experiments Provide?''}, Qu\'{\i}mica Nova, 17 (1994) 468-471
\bibitem{tsallis2} C.\ Tsallis, \textit{``Introduction to Nonextensive 
    Statistical Mechanics''}, Springer, 2009
\bibitem{cgk07}F.\ Clementi, M.\ Gallegati, G.\ Kaniadakis,
    \textit{``$\kappa$-Generalised Statistics in Personal Income
    Distribution''} Eur.\ Phys.\ J.\ B, 57 (2007) 187-193,
    arXiv:physics/0607293v2
\bibitem{cmgk08}F.\ Clementi, T.\ Di Matteo, M.\ Gallegati,
    G.\ Kaniadakis, \textit{``The $\kappa$-Generalised Distribution: a
    New Descriptive Model for the Size Distribution of Incomes''},
    Physica A, 387 (2008) 3201-3208, arXiv:0710.3645v4
\bibitem{cgk09}F.\ Clementi, M.\ Gallegati, G.\ Kaniadakis,
    \textit{``A $\kappa$-Generalized Statistical Mechanics Approach to
    Income Analysis''}, J.\ Stat.\ Mech., February (2009) P02037,
    arXiv:0902.0075v2
\bibitem{cgk12}F.\ Clementi, M.\ Gallegati, G.\ Kaniadakis,
    \textit{``A generalized statistical model for the size distribution
    of wealth''}, J.\ Stat.\ Mech., December (2012) P12006,
    arXiv:1209.4787v2
\bibitem{yamano} T.\ Yamano, \textit{``Some properties of $q$-logarithm
    and $q$-exponential functions in Tsallis statistics''}, Physica A 305
    (2002) 486-496
\bibitem{wilk} G.\ Wilk, Z.\ W{\l}odarczyk, \textit{``Tsallis
    Distribution Decorated with Log-Periodic Oscillation''},
    Entropy 17 (2015) 384-400, arXiv:1501.02608v1
\bibitem{huang} Y.\ Huang, H.\ Saleur, C.\ Sammis, D.\ Sornette,
    \textit{``Precursors, aftershocks, criticality and self-organized
    criticality''}, Europhys.\ Lett.\ 41 (1998) 43-48
\bibitem{sornette} D.\ Sornette, A.\ Johansen, J.-P.\ Bouchaud,
    \textit{``Stock market crashes, precursors and replicas''},
    J.\ Phys.\ I France 6 (1996) 167-175, arXiv:cond-mat/9510036v1 
\bibitem{vande1} N.\ Vandewalle, Ph.\ Boveroux, A.\ Minguet, M.\
    Ausloos, \textit{``The crash of October 1987 seen as a phase
    transition: amplitude and universality''}, Physica A 255 (1998)
    201-210
\bibitem{vande2} N.\ Vandewalle, M.\ Ausloos, Ph.\ Boveroux, A.\ Minguet,
    \textit{``How the financial crash of October 1987 could have been
    predicted''}, Eur.\ Phys. J.\ B 4 (1998) 139-141
\bibitem{wosnitza} J.H.\ Wosnitza, J.\ Leker, \textit{``Can log-periodic
    power-law structures arise from random fluctuations?''}, 
    Physica A 401 (2014) 228-250
\end{thebibliography}
\end{document}